1

# Dynamical Models of Stock Prices Based on Technical Trading Rules
# Part III: Application to Hong Kong Stocks

Li-Xin Wang

*Abstract*—In Part III of this study, we apply the price dynamical model with big buyers and big sellers developed in Part I of this paper to the daily closing prices of the top 20 banking and real estate stocks listed in the Hong Kong Stock Exchange. The basic idea is to estimate the strength parameters of the big buyers and the big sellers in the model and make buy/sell decisions based on these parameter estimates. We propose two trading strategies: (i) Follow-the-Big-Buyer which buys when big buyer begins to appear and there is no sign of big sellers, holds the stock as long as the big buyer is still there, and sells the stock once the big buyer disappears; and (ii) Ride-the-Mood which buys as soon as the big buyer strength begins to surpass the big seller strength, and sells the stock once the opposite happens. Based on the testing over 245 two-year intervals uniformly distributed across the seven years from 03-July-2007 to 02-July-2014 which includes a variety of scenarios, the net profits would increase 67% or 120% on average if an investor switched from the benchmark Buy-and-Hold strategy to the Follow-the-Big-Buyer or Ride-the-Mood strategies during this period, respectively.

## I. Introduction

In the real financial world, talking is cheap, and you must "show your muscle" through buying and selling (see for example [42]). Therefore, a good theory should lead to profitable trading strategies. The stock market is a battlefield among traders with different beliefs [8], [18], [22], [27], [32], [36], [42], and it is the winner's belief that determines the destiny of the stock prices[1]. Therefore, our basic trading philosophy is to follow the winners [47]. But, who are the winners? In this paper we concentrate on one type of traders who we believe are very likely to be the winners: the big buyers and the big sellers[2] --- the institutional investors (pension funds, mutual funds, hedge funds, money managers, investment banks, etc.) who manage large sums of money and often buy or sell a stock in large quantity. We make the following basic assumption:

**Assumption 1:** The stock prices are mainly determined by the actions of the institutional investors; we call them big buyers and big sellers.

To follow the big buyers/sellers, we must first find some way to detect them based on the noisy price data which is the only information we (small, individual investors) have. This seems to be a mission impossible, given the fact that the main task of these large investors is to keep their intentions as secret as possible by implementing a variety of sophisticated strategies to confuse or mislead other traders [20], [23], [25]. However, above all these ever-changing strategies [1], [2], [5], [12], [35], [37], one thing is almost certain: the big buyer (seller) must buy (sell) a large amount of the stock within a reasonable period of time with prices not too much higher (lower) than the market price at the time when the buying (selling) decision was made [6]. The main problem facing the big buyers and the big sellers is that the number of stocks available for selling and the money offered for purchasing the stocks at any time instant are very limited [15], consequently, the big buyers and the big sellers have no choice but to buy or sell the stock incrementally (little by little) over a period of time lasting for weeks or even months [30]. Statistically, these actions of the big buyers and the big sellers make the signed returns of the stocks a long memory process --- the autocorrelation function of the signed returns decays very slowly --- a well-known stylized fact in finance [9], [10]. And, these persistent buy or sell actions give us (the small investors) a chance to detect the presence of the big buyers and the big sellers and follow them up --- this is what we are going to do in this paper.

To do such detection, let's analyze in what situations the big buyers are most likely to buy and in what situations the big

---

[1] If there were no winners, the prices would be purely random [33] and the Efficient Market Hypothesis [14] would be true in this case.

[2] Of course there are other types of winners such as "the crowd" who have the so-called "swarm intelligence" [21], [43] and follow Keynes' Beauty Contest Theory (chapter 12 of [22]). Whoever the winners were, their existence would create some kinds of trends so that the markets could not be efficient in these situations.

sellers are most likely to sell. Consider the situation where the price is rising (the current price is above a moving average of the past prices), if the big buyers were still buying in this situation, the price would increase even further, resulting in higher cost for the big buyers (buy the stock with higher prices). Similarly, if the big sellers were still selling when the price is declining, the price would decline even further, causing a great increase of the cost for the big sellers (sell the stock at lower prices). It seems inevitable that the big buyers and the big sellers have to adapt (more or less) to the following trading heuristics:

**Big Buyer Heuristic:** For a big buyer, buy if the price of the stock is decreasing: the larger the decrease, the larger the buy order. Keep neutral when the price is increasing or moving horizontally.

**Big Seller Heuristic:** For a big seller, sell if the price of the stock is increasing: the stronger the increase, the larger the sell order. Keep neutral when the price is declining or moving horizontally.

which are the Heuristics 7 and 6 in Part I of this paper, respectively. So here we make another basic assumption:

**Assumption 2:** The big buyers (big sellers) use the above Big Buyer Heuristic (Big Seller Heuristic) in their real trading.

To further justify the Big Buyer and Big Seller Heuristics, we quote below what Walter Deemer[3] said in an interview by Andrew Lo (page 26 of [30]):

> When I went to work for Putnam Investments in 1970, in my very early days, I told one of the big fund managers that IBM had just broken out. "Fine, but I can't buy it," he said. "It's already broken out and moving, the price is rallying. I can't really buy it in size. So what I need you to do is tell me before it breaks out." So I went back to my office and spent quite a bit of time working on that, but I soon found out that when you're dealing with major institutions, managing large sums of money, you need to tell them that the only time they can really buy in quantity is during the decline, and the only time they can really sell in quantity is in a rally.

The last two lines of the quote above are the Big Buyer and Big Seller Heuristics.

Based on Assumptions 1 and 2, the basic idea of our approach is the following. First, using fuzzy systems theory [46], we convert the Big Buyer and Big Seller Heuristics into a price dynamical model that describes the stock price dynamics when the big buyers and the big sellers are the dominant forces in the market. Then, based on the real stock price data and using some parameter estimation algorithm developed in control theory [3], [17], we estimate the parameters in the price dynamical model that represent the strengths of the big buyers and the big sellers. Finally, two trading strategies --- Follow-the-Big-Buyer (FollowBB) and Ride-the-Mood (RideMood) --- are developed based on the estimated strength parameters. The basic idea of the FollowBB strategy is to buy the stock as soon as a big buyer is detected and at the same time no big seller is detected, and sell the stock when the big buyer stops buying. The basic idea of the RideMood strategy is to buy the stock as soon as the big buyer strength becomes larger than the big seller strength, and sell the stock when the opposite happens.

In what market conditions are the FollowBB and RideMood strategies likely to make money? In addition to Assumptions 1 and 2, we see from the basic ideas of the strategies that they are likely to make money when there is a dominant big buyer in the market. Indeed, if there are many big buyers and big sellers trading the stock around the same time, it will be difficult to predict who are going to win and consequently to determine the destiny of the stock prices. If everyone is big, then no one is big in comparison. We may make an analogy with boxing: if all fighters are heavyweights, it will be difficult to predict who are going to win the fights; however, if a heavyweight is fighting with a middleweight, we are much more certain about who will win. Taking the analogy back to stock trading, we make our third assumption as follows:

**Assumption 3:** For some stocks around some time periods, there is a dominant big buyer trading the stock. (Of course, we do not know which stocks and at what time periods the big buyer is trading; the task of our trading strategies is to find such stocks and such time periods.)

In what kind of stock markets and for what types of stocks is Assumption 3 most likely to be true? The answer is: the stock markets and the stocks which the "hot money" targets. We argue that the Hong Kong stock market is a good place where Assumptions 1 to 3 are very likely to be true. First, Hong Kong is one of the main financial centers in the world and all major financial firms have offices in Hong Kong, therefore we can confidently say that Assumption 1 is true for the Hong Kong stock market. Second, the Hong Kong Stock Exchange is a purely electronic limit order book market (no over-the-counter facilities) and the Hong Kong society has the tradition of strictly obeying the laws (under-the-table deals are rare, and law enforcement is strong), therefore large trades have to go directly to the limit order books and, consequently, Assumption 2 is most likely to be true. Third, because a) Hong Kong is the top free market in the world for many years (there is no restriction on capital in-flow and out-flow), b) the taxes in Hong Kong are low, and c) most big companies in China are co-listed in the Hong Kong Stock Exchange so that funds with "China concept" can go to Hong Kong to make the deal (the capital markets in Mainland China are not freely open to foreign investors), the Hong Kong stock market is an attractive place for the "hot money" (e.g., Soros's Quanta Fund has an office in Hong Kong) and, consequently, the market condition of Assumption 3 is expected to appear very frequently in the Hong Kong stock market. Therefore, we use Hong Kong stock market as the first stop to test our trading strategies.

---

[3] Walter Deemer is the featured technical analyst in Dean LeBaron's book *Dean LeBaron's Treasury of Investment Wisdom*, joining such luminaries as John Bogle, Peter Lynch, and George Soros as the chosen gurus in their fields (quote from page 25 of [30]).

The rest of the paper is organized as follows. In Section II, we will reproduce the price dynamical model from the Big Buyer and Big Seller Heuristics and explain the basic ideas of the FollowBB and RideMood strategies. In Section III, we will use the recursive least-squares algorithm with exponential forgetting to estimate the model parameters and simulate the algorithm for our strongly nonlinear price dynamical model to get some feeling of the performance of the algorithm in different noisy conditions. In Section IV, we will present the details of the FollowBB and RideMood strategies, and we will also present the benchmark Buy-and-Hold (Buy&Hold) and the classic Trend-Following (TrendFL) strategies which will be used to compare with our FollowBB and RideMood strategies for real stock applications in the later sections. In Section V, we will apply the four trading strategies --- FollowBB, RideMood, TrendFL, and Buy&Hold --- to the top 20 banking and real estate stocks in Hong Kong Stock Exchange for the recent seven-year period from July 3, 2007 to July 2, 2014 with daily closing data. In Section VI, we will combine the 20 stocks into a simple portfolio scheme to compare the overall performance of the trading strategies. In Section VII, we will show the details of the buy-sell cycles of our FollowBB and RideMood strategies for three recent full years from January 3, 2011 to December 31, 2013. Finally, Section VIII concludes the paper.

## II. PRICE DYNAMIC MODEL AND BASIC IDEAS OF THE TRADING STRATEGIES

Using the basic tools in fuzzy systems theory [46] (fuzzy sets, fuzzy IF-THEN rules, fuzzy systems, etc.), we transformed the Big Buyer and Big Seller Heuristics (the Heuristics 7 and 6 in Part I of this paper) into the following price dynamical model in Part I of this paper:

$$\ln(p_{t+1}) = \ln(p_t) + a_6(t)\, ed_6\big(x_t^{(1,n)}\big) + a_7(t)\, ed_7\big(x_t^{(1,n)}\big) + \varepsilon(t) \quad (1)$$

where $p_t$ is the price of the stock at time $t$, $ed_6$ ($ed_7$) is the excess demand of the big seller (buyer), $a_6$ ($a_7$) is the strength parameter of the big seller (buyer), $\varepsilon(t)$ is the price impact of other traders except the big seller and the big buyer, and $x_t^{(1,n)}$ is the log-ratio of price $p_t$ to its $n$-day moving average:

$$x_t^{(1,n)} = \ln\left(p_t \Big/ \tfrac{1}{n}\sum_{i=0}^{n-1} p_{t-i}\right) \quad (2)$$

which is the basic variable characterizing the trend of the price series. The excess demands $ed_6$ and $ed_7$ are fuzzy systems constructed from the Big Seller Heuristic and the Big Buyer Heuristic, respectively, and their detailed formulas are (from Part I of this paper):

$$ed_6(x_t^{(1,n)}) = \frac{-0.1\,\mu_{PS}(x_t^{(1,n)}) - 0.2\,\mu_{PM}(x_t^{(1,n)}) - 0.4\,\mu_{PL}(x_t^{(1,n)})}{\mu_{PS}(x_t^{(1,n)}) + \mu_{PM}(x_t^{(1,n)}) + \mu_{PL}(x_t^{(1,n)}) + \mu_{AZ}(x_t^{(1,n)})} \quad (3)$$

$$ed_7(x_t^{(1,n)}) = \frac{0.1\,\mu_{NS}(x_t^{(1,n)}) + 0.2\,\mu_{NM}(x_t^{(1,n)}) + 0.4\,\mu_{NL}(x_t^{(1,n)})}{\mu_{NS}(x_t^{(1,n)}) + \mu_{NM}(x_t^{(1,n)}) + \mu_{NL}(x_t^{(1,n)}) + \mu_{AZ}(x_t^{(1,n)})} \quad (4)$$

where the membership functions µ's are given in Fig. 1 of Part I of this paper. Substituting these membership functions into (3) and (4) yields:

$$ed_6\big(x_t^{(1,n)}\big) = \begin{cases} 0 & x_t^{(1,n)} \le 0 \\ -\dfrac{0.1 x_t^{(1,n)}}{w} & 0 \le x_t^{(1,n)} \le 2w \\ -\dfrac{0.2 x_t^{(1,n)}}{w} + 0.2 & 2w \le x_t^{(1,n)} \le 3w \\ -0.4 & 3w \le x_t^{(1,n)} \end{cases} \quad (5)$$

$$ed_7\big(x_t^{(1,n)}\big) = \begin{cases} 0.4 & x_t^{(1,n)} \le -3w \\ -\dfrac{0.2 x_t^{(1,n)}}{w} - 0.2 & -3w \le x_t^{(1,n)} \le -2w \\ -\dfrac{0.1 x_t^{(1,n)}}{w} & -2w \le x_t^{(1,n)} \le 0 \\ 0 & 0 \le x_t^{(1,n)} \end{cases} \quad (6)$$

which are plotted in Fig. 1. Notice from (5) and (6) or Fig. 1 that at any value of $x_t^{(1,n)}$, either $ed_6(x_t^{(1,n)})$ or $ed_7(x_t^{(1,n)})$ is zero. More specifically, when the price is rising ($x_t^{(1,n)} > 0$), the big buyer excess demand $ed_7(x_t^{(1,n)})$ is zero; when the price is declining ($x_t^{(1,n)} < 0$), the big seller excess demand $ed_6(x_t^{(1,n)})$ is zero.

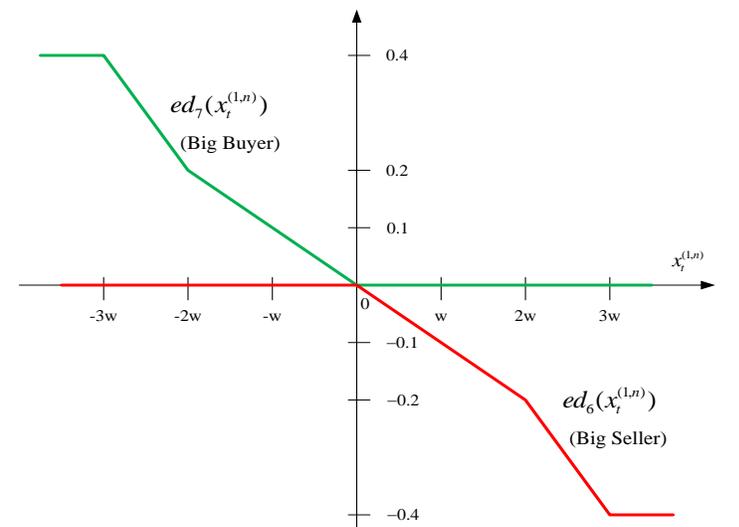

Fig. 1: Excess demand functions $ed_6(x_t^{(1,n)})$ (big seller) and $ed_7(x_t^{(1,n)})$ (big buyer).

The first step of our trading strategies is to estimate the strength parameters $a_6(t)$ and $a_7(t)$ based on the price data up to the current time *t*. Then, the trading strategies are developed based on the following three simple arguments:
- Positive $a_6(t)$ implies the existence of big sellers;
- Positive $a_7(t)$ implies the existence of big buyers;
- $a_7(t) > a_6(t)$ indicates that the price is in a rising mode, and $a_7(t) < a_6(t)$ means that the price is in a declining mode.

From the first two arguments above we propose a trading strategy called "Follow-the-Big-Buyer", and based on the third argument above we propose a trading strategy called "Ride-the-Mood"; the basic ideas of these two strategies are as follows:

- **Follow-the-Big-Buyer (FollowBB):** Buy the stock once $a_7(t)$ becomes positive and $a_6(t)$ is negative (which means there are big buyers coming and there are no big sellers); hold the stock as long as $a_7(t)$ is still positive (which means do not sell the stock as long as the big buyers are still buying, no matter what happens to the price or whether there are big sellers; stand firm with your big buyers); sell the stock once $a_7(t)$ becomes negative (quit if your big buyers stop buying); the buy-sell round trip is completed and watch out for the next cycle.
- **Ride-the-Mood (RideMood):** Buy the stock once $a_7(t) - a_6(t)$ becomes positive (which means the big buyers are gaining an upper hand over the big sellers); hold the stock as long as $a_7(t) - a_6(t)$ is still positive; sell the stock once $a_7(t) - a_6(t)$ becomes negative (which means the big sellers become strong than the big buyers); the buy-sell round trip is completed and watch out for the next cycle.

In Section IV, we will give the detailed flow-charts of the FollowBB and RideMood strategies that require the estimates of the strength parameters $a_6(t)$ and $a_7(t)$, which are provided by the parameter estimation algorithm in the next section.

### III. PARAMETER ESTIMATION ALGORITHM

Given the price data $\{p_0, p_1, \ldots, p_t, p_{t+1}\}$, our goal is to estimate the strength parameters $a_6(t)$ and $a_7(t)$ in model (1) based on this information set. Let $r_{t+1} = \ln(p_{t+1}/p_t)$ be the return and define

$$a_t = (a_6(t), a_7(t))^T \qquad (7)$$

$$ed_t = \left(ed_6(x_t^{(1,n)}), ed_7(x_t^{(1,n)})\right)^T \qquad (8)$$

then the price dynamical equation (1) becomes

$$r_{t+1} = ed_t^T a_t + \varepsilon(t) \qquad (9)$$

We can reasonably assume that the strength parameters $a_t = (a_6(t), a_7(t))^T$ of the big buyer/seller are slowly time-varying, because a large buy/sell order has to be cut into small pieces and implemented incrementally over a long period of time due to the small liquidity available in the order book. A good method to estimate slowly time-varying parameters is the standard Recursive Lease Squares Algorithm with Exponential Forgetting which minimizes the weighted summation of error:

$$E_{t+1}(a) = \sum_{i=1}^{t} \lambda^{t-i} (r_{i+1} - ed_i^T a)^2 \qquad (10)$$

to obtain the estimate of $a_t$, denoted as $\hat{a}_t$, through the following recursive computations (see, e.g., page 53 of [3]):

$$\hat{a}_t = \hat{a}_{t-1} + K_t(r_{t+1} - ed_t^T \hat{a}_{t-1}) \qquad (11)$$

$$K_t = \frac{P_{t-1} ed_t}{(ed_t^T P_{t-1} ed_t + \lambda)} \qquad (12)$$

$$P_t = (I - K_t ed_t^T) P_{t-1} / \lambda \qquad (13)$$

where the initial $\hat{a}_0 = 0$, $P_0 = \gamma I$ for some large $\gamma$, and $\lambda \in (0,1)$ is a forgetting factor to put more weights on recent data as in (10). Certainly, other parameter estimation algorithms [17], [19], [45] can be used to estimate the $a_t$.

Before applying the algorithm to real stock data to identify big buyers and big sellers in the following sections, we perform simulations to get some feeling about the performance of the parameter estimation algorithm. Let the price $p_t$ be generated by model (1) with *n*=3, *w*=0.01, initial prices $p_{-2} = p_{-1} = p_0 = 10$, $a_7(t)$ and $-a_6(t)$ given as the blue lines in the top sub-figures of Figs. 2 to 4, and $\varepsilon(t)$ being an i.i.d. zero-mean Gaussian random process with standard deviation $\sigma$. Fig. 2 shows a simulation run of model (1) and the parameter estimation algorithm (11)-(13) with $\sigma = 0.02$, $\lambda = 0.95$ and $\gamma = 10$, where the bottom sub-figure is the price series $p_t$ and the upper sub-figure plots the true parameters $a_7(t)$ and $-a_6(t)$ (blue lines) and their estimates $\hat{a}_7(t)$ and $-\hat{a}_6(t)$ (red lines). We see from Fig. 2 that the parameter estimates $\hat{a}_7(t)$ and $-\hat{a}_6(t)$ can in general catch up with the changes of the true parameters $a_7(t)$ and $-a_6(t)$, but the estimates $\hat{a}_7(t)$ and $-\hat{a}_6(t)$ are noisy and delayed.

To judge whether the estimates $\hat{a}_7(t)$ and $-\hat{a}_6(t)$ in Fig. 2 are good or bad, we need to determine the signal-to-noise ratio of the prices $p_t$ used in Fig. 2. Since in model (1) the signal term is $a_6(t)ed_6(x_t^{(1,n)}) + a_7(t)ed_7(x_t^{(1,n)})$ and the noise term is $\varepsilon(t)$, we define the signal-to-noise ratio as follows:

$$\frac{S}{N} = \frac{\left(\frac{1}{N}\sum_{t=1}^{N}\left(a_6(t)\,ed_6(x_t^{(1,n)}) + a_7(t)\,ed_7(x_t^{(1,n)})\right)^2\right)^{\frac{1}{2}}}{\left(\frac{1}{N}\sum_{t=1}^{N}(\varepsilon(t))^2\right)^{\frac{1}{2}}} \qquad (14)$$

For the price $p_t$ in Fig. 2, the signal-to-noise ratio is 1.0518, which means the price impact of the big seller plus the big buyer is roughly equal to the summation of the price impacts of the rest of the traders.

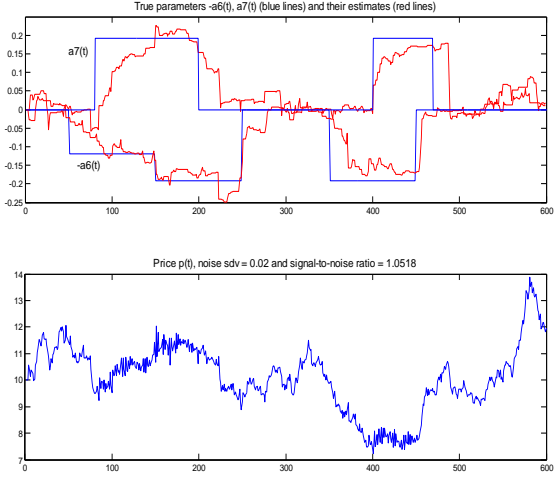

Fig. 2: Simulation of the parameter estimation algorithm for the case of signal-to-noise = 1.0518. Top: true $a_7(t)$ and $-a_6(t)$ (blue lines) and their estimates (red lines). Bottom: the price $p_t$.

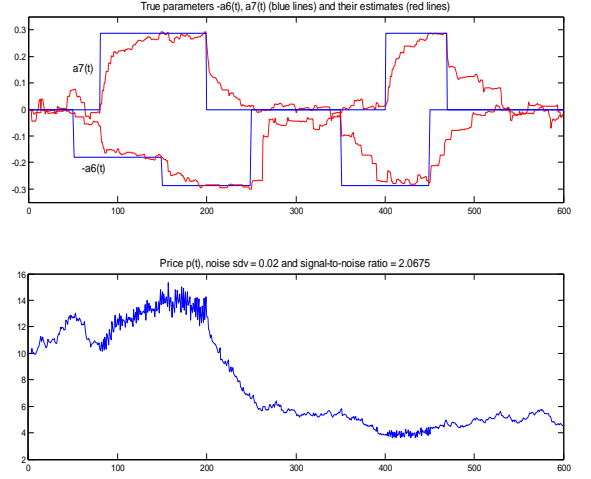

Fig. 3: Simulation of the parameter estimation algorithm for the case of signal-to-noise = 2.0675. Top: true $a_7(t)$ and $-a_6(t)$ (blue lines) and their estimates (red lines). Bottom: the price $p_t$.

To see the performance of the estimation algorithm for more signal-to-noise cases, we show in Figs. 3 and 4 the simulation results with signal-to-noise ratio equal to 2.0675 and 0.50156, respectively. That is, for the price $p_t$ in Fig. 3 (Fig. 4), the price impact of the big seller plus the big buyer is about twice (half) as strong as the summation of the price impacts of the rest of the traders. From Figs. 2 to 4 we conclude that:

- The stronger the strength of the big buyer/seller with respect to the rest of the traders, the better the estimates of the strength parameters of the big buyer/seller;
- In the cases where the big buyer/seller is as strong as or stronger than the rest of the traders, we have good chance to identify the existence of the big buyer/seller in a timely fashion and to estimate their strengths correctly;
- Even in the case where the strength of the big buyer/seller is only about half of the other traders, we can still manage to detect the existence of the big buyer/seller (with longer delay).

Since big buyer/seller by definition should be stronger than other traders (otherwise they are not big), the conclusions above give us confidence to apply the estimation algorithm to real stock data to detect the real hidden big buyers/sellers and follow them up.

## IV. THE TRADING STRATEGIES: FOLLOWBB, RIDEMOOD, TRENDFL AND BUY&HOLD

The basic ideas of the Follow-the-Big-Buyer (FollowBB) and the Ride-the-Mood (RideMood) strategies were discussed in Section II, and the detailed algorithms are now given in Figs. 5 and 6 for FollowBB and RideMood, respectively, where

$$\bar{a}_6(t,m) = \frac{1}{m}\sum_{i=0}^{m-1}\hat{a}_6(t-i), \quad \bar{a}_7(t,m) = \frac{1}{m}\sum_{i=0}^{m-1}\hat{a}_7(t-i) \quad (15)$$

are the $m$-day moving averages of the estimated parameters $\hat{a}_6(t)$ and $\hat{a}_7(t)$ which are obtained from the parameter estimation algorithm (11) to (13) ($m=3$ for FollowBB and $m=5$ for RideMood). We use moving averages of the parameter estimates because we see from the simulation results in Figs. 2-4 that the parameter estimates were noisy, and moving averages can smooth out the noise. The positive side of using moving averages is that false alarm rate can be reduced due to the reduction of noise, whereas its negative effect is the increase of delays in detecting the arrival and depart of the big buyer/seller. Notice from Figs. 5 and 6 that for a single stock using these strategies, there are only two status: full-cash or full-stock, no mixed cash/stock status.

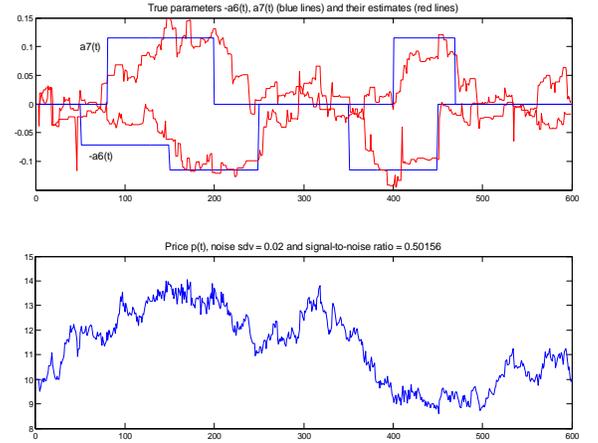

Fig. 4: Simulation of the parameter estimation algorithm for the case of signal-to-noise = 0.5015. Top: true $a_7(t)$ and $-a_6(t)$ (blue lines) and their estimates (red lines). Bottom: the price $p_t$.

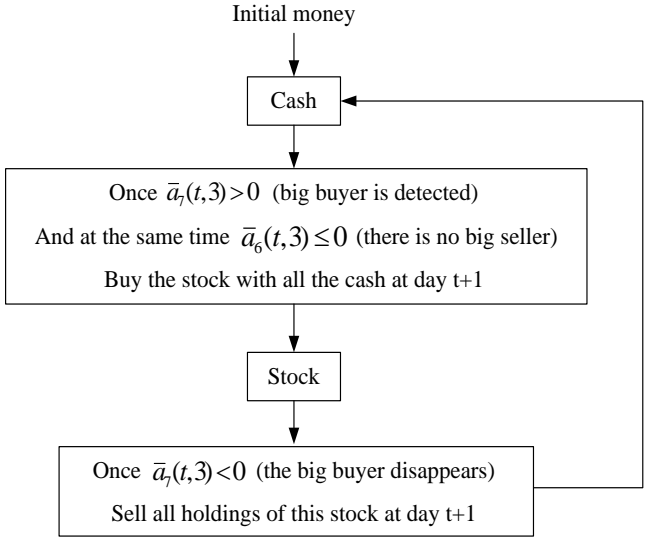

Fig. 5: The Follow-the-Big-Buyer (FollowBB) strategy.

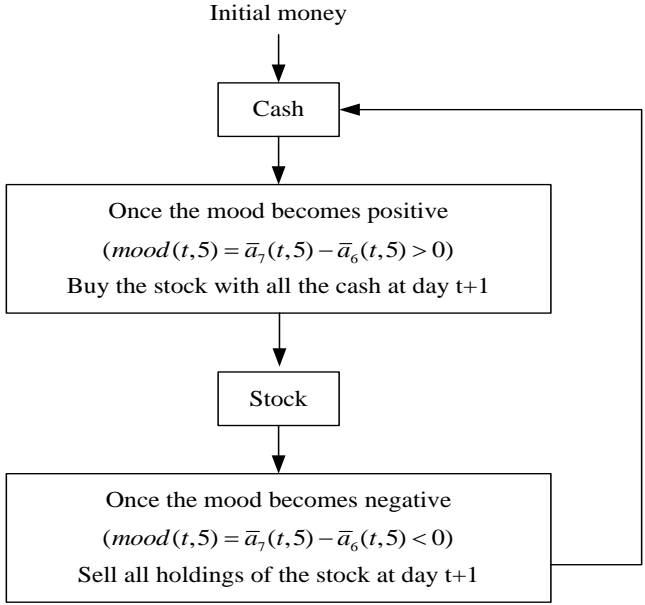

Fig. 6: The Ride-the-Mood (RideMood) strategy.

To show the outstanding performance of the FollowBB and RideMood strategies, we will compare them with other three popular trading strategies: the benchmark Buy-and-Hold strategy [33], the classic Trend-Following strategy of technical traders [4], [24], and the Modern Portfolio Theory's favorite Index Fund (the Index return) [13]. The Buy-and-Hold strategy is as follows:

- **Buy-and-Hold (Buy&Hold):** Buy the stock in the first day of the investment interval with all the cash allocated to this stock, hold the stocks until the last day of the investment interval and at that time sell all the holdings of this stock.

which is the standard benchmark for comparing different investment schemes, because it is often claimed in the academic literature that no strategy can outperform the Buy-and-Hold strategy consistently (see, e.g., [33]). In all of our test scenarios in this paper, we will show the returns of the Buy-and-Hold strategy as the benchmark for comparison. The basic idea of the classic Trend-Following strategy is [24]:

- **Trend-Following (TrendFL):** Buy when a shorter moving-average is across a longer moving-average from below (a rising mode starts), and sell once the shorter moving-average is back below the longer moving-average (the rising mode ends).

Fig. 7 gives the details of the TrendFL strategy, where 5-day (a trading week) and 60-day (roughly a trading season) are chosen for the shorter and longer moving averages, respectively.

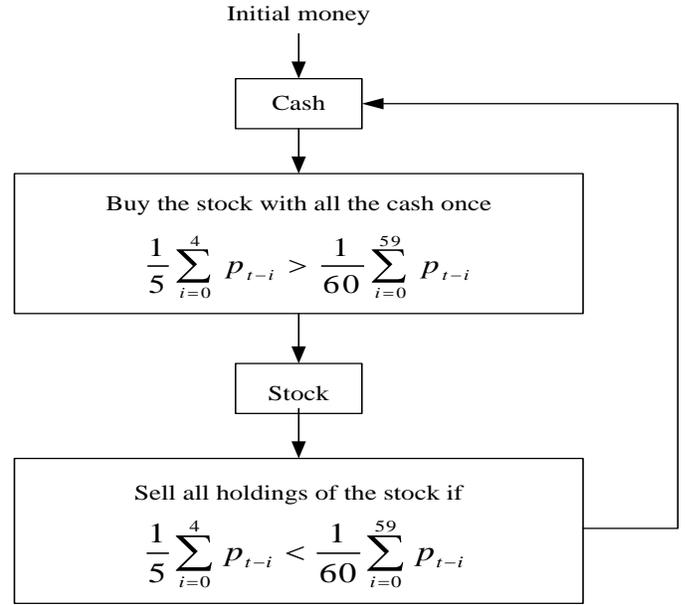

Fig. 7: The Trend-Following (TrendFL) strategy.

V. APPLICATION TO HONG KONG STOCKS

We now apply the four trading strategies --- FollowBB, RideMood, TrendFL and Buy&Hold --- to the daily closing prices of the top 20 banking and real estate stocks listed in the Hong Kong Stock Exchange (Table 1). We choose these top banking and real estate stocks because they are usually the targets of the "hot money" due to their representativeness to the economical conditions in Hong Kong and China. We will use the daily closing price data[4] of these 20 stocks during the recent seven-year period from July 3, 2007 to July 2, 2014 to test the trading strategies. Since the market conditions were changing wildly during this period (the top sub-figure of Fig. 8 shows the Hang Seng Index (HSI) daily closings during this period which includes: 1) the strong rise in late 2007, 2) the big decline due to the 2008 financial crisis, 3) the recoverary after the panic, and 4) the "ordinary" years from 2010 to 2014), it gives us a good chance to test the trading strategies in different market conditions.

---

[4] All stock price data used in this paper were downloaded from http://finance.yahoo.com and were adjusted for dividends and splits.

We choose two years as the length for performance evaluation, and test the trading strategies for a large number of two-year periods uniformly distributed over these seven years. Specifically, the bottom sub-figure of Fig. 8 illustrates the scheme: each test interval has 492 trading days (roughly two years), the first test interval starts from 03-07-2007, the second test interval starts 5 trading days later, the third starts another 5 trading days later, ..., and the 245th test interval ends around 02-07-2014 so that the whole seven years from 03-07-2007 to 02-07-2014 are covered. For each test interval, we ran the FollowBB in Fig. 5, the RideMood in Fig. 6, the TrendFL in Fig. 7 and the Buy&Hold for each of the 20 stocks in Table 1 using their daily closing prices as the $p_t$ in the parameter estimation algorithm (11) to (13) with $\lambda = 0.95$, $\gamma = 10$, $n=3$ and $w=0.01$ which are determined based on a number of trial-and-error experiments[5].

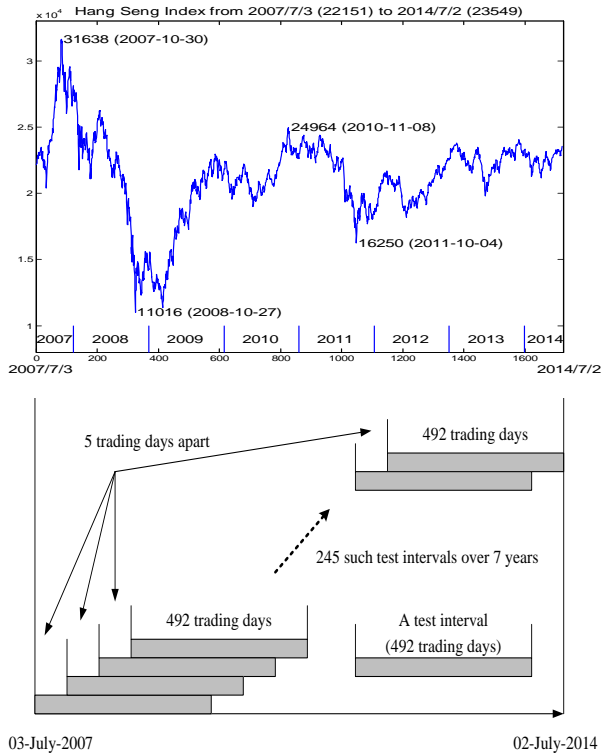

Fig. 8: Top: The Hang Seng Index (HSI) from 03-July-2007 (22151) to 02-July-2014 (23549). Bottom: The test intervals for the trading strategies: 492 trading days (roughly two years) per test interval, and 245 such test intervals in sequence with five trading days apart.

Let $p_{i,j,k}^{buy}$ ($p_{i,j,k}^{sell}$) be the buy (sell) price of the $k$'th buy-sell cycle (see Figs. 5 to 7) for stock $i$ ($i=1$ is HK0005, $i=2$ is HK0939, ..., $i=20$ is HK0267) in the $j$'th test interval in Fig. 8 ( $j=1,2, ..., 245$) and $N_{i,j}$ be the number of such buy-sell cycles, then the annual return of the trading strategy (FollowBB, RideMood, TrendFL or Buy&Hold) for stock $i$ over test interval $j$, after deducting the transaction costs, is

$$ar(i,j) = \frac{1}{2}\left(\prod_{k=1}^{N_{i,j}} \frac{p_{i,j,k}^{sell}}{p_{i,j,k}^{buy}}\right)(1 - 0.316\%)^{N_{i,j}} \quad (16)$$

where 0.316% is the transaction cost for one buy-sell cycle[6], and the $\frac{1}{2}$ is due to the fact that each test interval is two years and we are computing the annual returns.

Figs. 9.1 to 9.5 plot $ar(i,j)$ as function of $j$ for each of the 20 stocks ($i=1$ to 20) using the four strategies: FollowBB (green), RideMood (red), TrendFL (yellow) and Buy&Hold (blue). Also plotted in Figs. 9.1 to 9.5 is the Hang Seng Index annual returns (black) over the 245 test intervals in Fig. 8. From Figs. 9.1 to 9.5 we see that the annual returns change wildly over the 245 test intervals.

To get a summary of the overall performance of a trading strategy for a stock, define the *average annual return* for stock $i$ as:

$$\overline{ar}(i) = \frac{1}{245}\sum_{j=1}^{245} ar(i,j) \quad (17)$$

and its standard deviation:

$$sdv(i) = \left(\frac{1}{245}\sum_{j=1}^{245}\left(ar(i,j) - \overline{ar}(i)\right)^2\right)^{\frac{1}{2}} \quad (18)$$

Table 1 gives the $\overline{ar}(i)$'s, the $sdv(i)$'s and the Sharpe ratios $\overline{ar}(i)/sdv(i)$ (we assume the risk free rate is zero (taking cash back home), which is almost exactly true in Hong Kong after the 2008 financial crisis) of the four trading strategies for the 20 stocks in Table 1. The last item in Table 1 gives the average annual return of HSI over the 245 test intervals, the standard deviation and the Sharpe ratio.

From Figs. 9.1 to 9.5 and Table 1 we see that the performance of the trading strategies for different stocks changes wildly, implying that each stock has its own "personality". To evaluate the overall performance of the trading strategies, we combine the 20 stocks into a portfolio in the next section.

---

[5] This parameter setting cannot be optimal for all stocks. In real trading, some optimization procedures were very helpful to determine the most suitable parameter setting for each stock. In this scientific paper, we simply use this same parameter setting for all the stocks to show that even with this non-optimal parameter setting, our FollowBB and RideMood can still outperform the benchmark Buy&Hold and the classic TrendFL.

[6] The cost of a buy-sell cycle is computed as follows: Brokerage fee 0.1% of transaction amount for selling only (on promotion); Stamp duty 0.1% of transaction amount for both buying and selling; Transaction levy 0.003% of transaction amount for both buying and selling; and Trading fee 0.005% of transaction amount for both buying and selling. Other fees are ignorable comparing with the fees above. So the average cost per buy-sell cycle is: $(0.1\% + 0.1\% \times 2 + 0.003\% \times 2 + 0.005\% \times 2) = 0.316\%$.

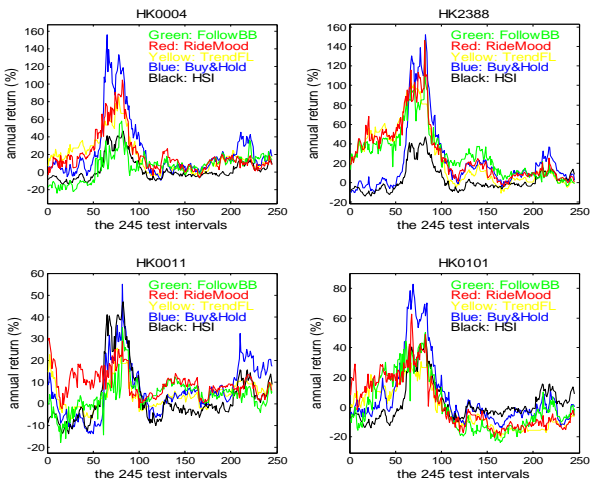

Fig. 9.1: Annual returns of FollowBB (green), RideMood (red), TrendFL (yellow), Buy&Hold (blue) and HSI (black) over the 245 test intervals in Fig. 8 for stocks HK0005, HK0939, HK1398 and HK3988.

Fig. 9.2: Annual returns of FollowBB (green), RideMood (red), TrendFL (yellow), Buy&Hold (blue) and HSI (black) over the 245 test intervals in Fig. 8 for stocks HK2628, HK0001, HK0016 and HK0388.

Fig. 9.3: Annual returns of FollowBB (green), RideMood (red), TrendFL (yellow), Buy&Hold (blue) and HSI (black) over the 245 test intervals in Fig. 8 for stocks HK0004, HK2388, HK0011 and HK0101.

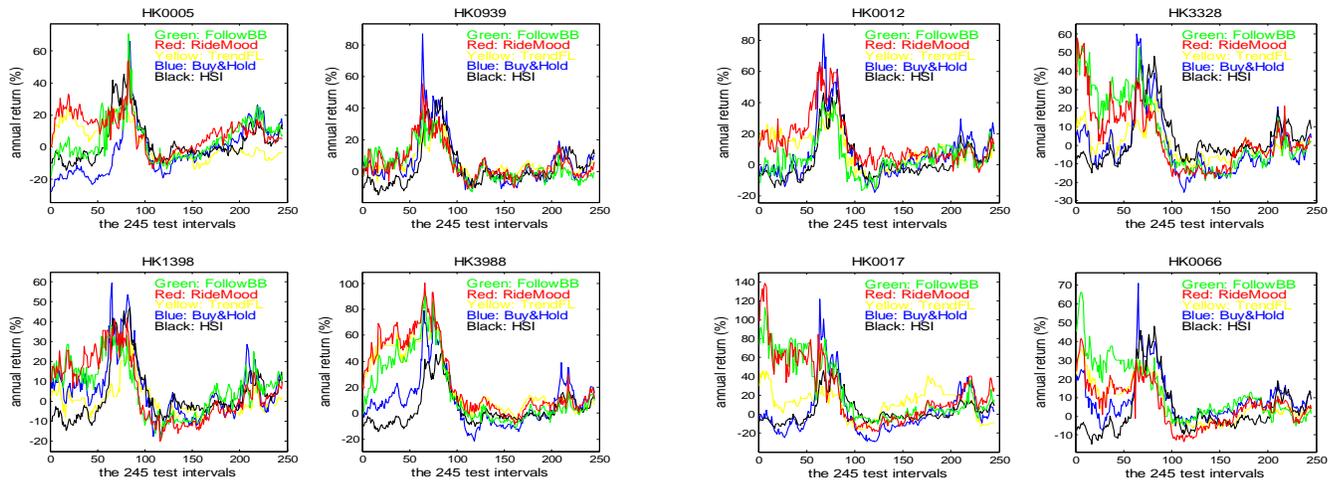

Fig. 9.4: Annual returns of FollowBB (green), RideMood (red), TrendFL (yellow), Buy&Hold (blue) and HSI (black) over the 245 test intervals in Fig. 8 for stocks HK0012, HK3328, HK0017 and HK0066.

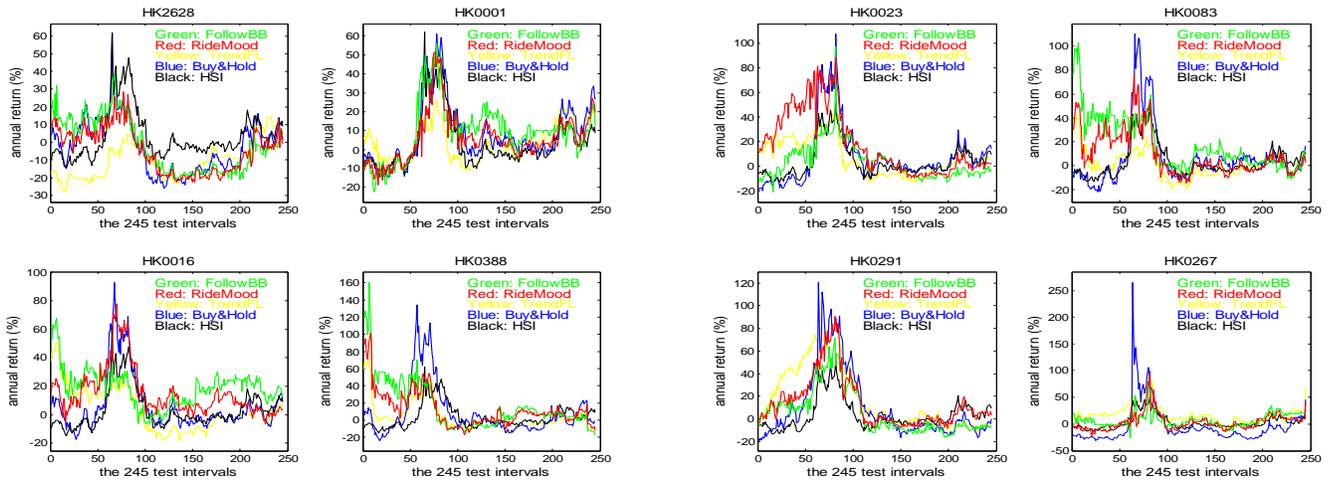

Fig. 9.5: Annual returns of FollowBB (green), RideMood (red), TrendFL (yellow), Buy&Hold (blue) and HSI (black) over the 245 test intervals in Fig. 8 for stocks HK0023, HK0083, HK0291 and HK0267.

## VI. PORTFOLIO PERFORMANCE

We propose a very simple portfolio scheme: Distribute the initial money to the 20 stocks according to their weights in HSI and run the trading strategy (FollowBB, RideMood, TrendFL or Buy&Hold) independently for each stock until the closure of the portfolio. Fig. 10 illustrates this scheme for an investment interval, where $w(i)$ is the weight of the $i$'th stock in Table 1 in HSI ($w(1) = 15\%$, $w(2) = 7.46\%$, ..., $w(20) = 0.22\%$). We see from Fig. 10 that once the initial money is distributed to the 20 stocks, no money will be redistributed among the stocks; that is, the investment for each stock is done independently. Hence, if a stock is doing good, we keep the profit within this stock for the subsequent buy/sell cycles; if a stock is doing badly, it will has less cash for its future investment.

Table 1: Average annual returns $\overline{ar}(i)$, standard deviations $sdv(i)$ and Sharpe ratios $\overline{ar}(i)/sdv(i)$ of the 20 stocks over the 245 test intervals in Fig. 8 using FollowBB, RideMood, TrendFL and Buy&Hold. The returns are net returns after deducting the transaction costs.

| Stock (1-7) | Trading strategy | Average Annual Return (%) ±standard deviation (%); Sharpe ratio | Stock (8-14) | Trading strategy | Average Annual Return (%) ±standard deviation (%); Sharpe ratio | Stock (15-20) | Trading strategy | Average Annual Return (%) ±standard deviation (%); Sharpe ratio |
|---|---|---|---|---|---|---|---|---|
| HK0005 HSBC Holdings plc | FollowBB RideMood TrendFL Buy&Hold | 4.78 ±12.60; 0.37<br>**9.51** ±11.06; 0.85<br>2.43 ±10.56; 0.23<br>-1.82 ±15.56; -0.11 | HK0388 Hong Kong Exc. & Cle. | FollowBB RideMood TrendFL Buy&Hold | **16.68** ±29.12; 0.57<br>12.59 ±21.58; 0.58<br>4.30 ±16.72; 0.25<br>8.75 ±29.25; 0.29 | HK0017 New World Development | FollowBB RideMood TrendFL Buy&Hold | **24.12** ±31.31; 0.77<br>24.17 ±34.20; 0.70<br>11.79 ±14.89; 0.79<br>3.12 ±17.23; 0.18 |
| HK0939 China Constr.Bank | FollowBB RideMood TrendFL Buy&Hold | 4.61 ±10.61; 0.43<br>6.00 ±11.56; 0.51<br>5.27 ±8.10; 0.65<br>**6.17** ±15.23; 0.40 | HK0004 The Wharf (Holdings) Ltd | FollowBB RideMood TrendFL Buy&Hold | 7.00 ±14.35; 0.48<br>20.07 ±21.90; 0.91<br>19.07 ±19.18; 0.99<br>**24.74** ±33.73; 0.73 | HK0066 MTR Corporation | FollowBB RideMood TrendFL Buy&Hold | **12.74** ±15.05; 0.84<br>5.87 ±11.41; 0.51<br>7.04 ±11.07; 0.63<br>9.17 ±11.60; 0.79 |
| HK1398 Ind. & Com. Bank of China | FollowBB RideMood TrendFL Buy&Hold | **6.52** ±12.97; 0.50<br>5.46 ±15.41; 0.35<br>0.45 ±7.47; 0.06<br>6.18 ±9.42; 0.65 | HK2238 BOC Hong Kong | FollowBB RideMood TrendFL Buy&Hold | **29.70** ±25.13; 1.18<br>30.00 ±29.75; 1.00<br>25.99 ±30.31; 0.85<br>22.34 ±34.37; 0.65 | HK0023 The Bank of East Asia | FollowBB RideMood TrendFL Buy&Hold | 3.52 ±14.34; 0.24<br>**18.48** ±26.00; 0.71<br>7.26 ±14.34; 0.50<br>8.24 ±23.90; 0.34 |
| HK3988 Bank of China Ltd | FollowBB RideMood TrendFL Buy&Hold | 17.18 ±24.67; 0.69<br>**25.93** ±27.64; 0.93<br>23.12 ±22.92; 1.01<br>9.48 ±20.10; 0.47 | HK0011 Hang Seng Bank | FollowBB RideMood TrendFL Buy&Hold | 3.97 ±7.03; 0.56<br>**9.44** ±6.17; 1.52<br>4.31 ±6.62; 0.65<br>6.41 ±6.95; 0.92 | HK0083 Sino Land Company Ltd | FollowBB RideMood TrendFL Buy&Hold | **18.34** ±22.29; 0.82<br>9.57 ±18.43; 0.51<br>1.08 ±10.98; 0.09<br>6.78 ±24.35; 0.27 |
| HK2628 China Life Insurance Co | FollowBB RideMood TrendFL Buy&Hold | **-1.76** ±14.83; -0.11<br>-4.33 ±12.51; -0.34<br>-10.15 ±10.5; -0.96<br>-2.70 ±14.57; -0.18 | HK0101 Hang Lung PPT | FollowBB RideMood TrendFL Buy&Hold | 0.93 ±18.14; 0.05<br>2.22 ±17.59; 0.12<br>-1.34 ±14.12; -0.09<br>**7.47** ±22.22; 0.33 | HK0291 China Resources Enterprise Ltd | FollowBB RideMood TrendFL Buy&Hold | 5.81 ±19.73; 0.29<br>14.05 ±22.62; 0.62<br>**19.20** ±28.16; 0.68<br>10.58 ±29.45; 0.35 |
| HK0001 Cheung Kong (Holdings) Ltd | FollowBB RideMood TrendFL Buy&Hold | **9.41** ±15.13; 0.62<br>6.25 ±12.87; 0.48<br>3.79 ±9.43; 0.40<br>8.44 ±15.90; 0.53 | HK0012 Henderson Land Dev. | FollowBB RideMood TrendFL Buy&Hold | 4.52 ±11.19; 0.40<br>**14.92** ±15.83; 0.94<br>9.02 ±11.46; 0.78<br>7.74 ±18.28; 0.42 | HK0267 CITIC Pacific Ltd | FollowBB RideMood TrendFL Buy&Hold | 7.35 ±12.91; 0.56<br>3.69 ±16.34; 0.22<br>**18.34** ±17.49; 1.04<br>-3.68 ±35.43; -0.10 |
| HK0016 Sun Hung Kai Properties Ltd | FollowBB RideMood TrendFL Buy&Hold | **18.62** ±13.11; 1.42<br>15.18 ±16.88; 0.89<br>3.97 ±14.98; 0.26<br>6.44 ±15.18; 0.42 | HK3328 Bank of Communic. | FollowBB RideMood TrendFL Buy&Hold | **6.07** ±19.84; 0.30<br>3.52 ±16.34; 0.21<br>0.54 ±8.22; 0.06<br>-1.09 ±15.76; -0.06 | HSI | | **3.27** ±13.35; 0.24 |

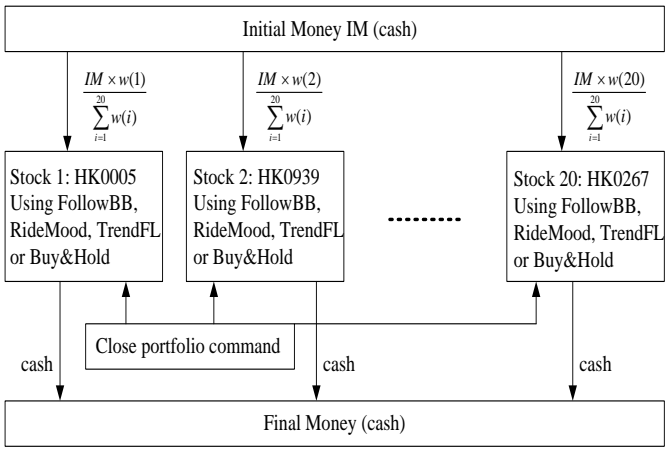

Fig. 10: Portfolio scheme for a test interval in Fig. 8.

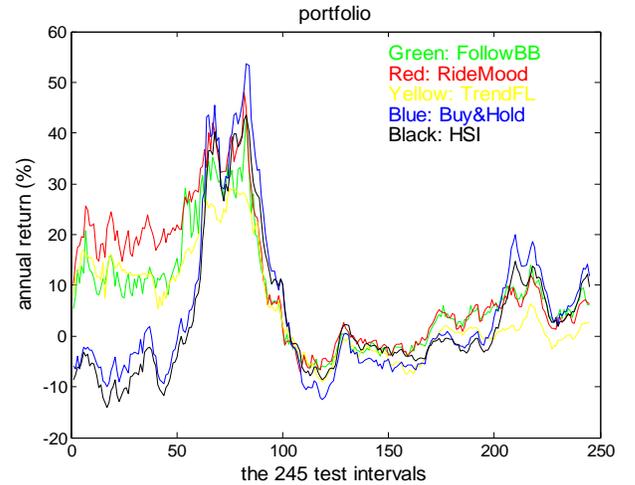

Fig. 11: Annual returns of HSI (black) and the portfolio using FollowBB (green), RideMood (red), TrendFL (yellow) and Buy&Hold (blue) over the 245 test intervals in Fig. 8.

Fig. 11 plots the annual returns of the portfolio using the four strategies: FollowBB (green), RideMood (red), TrendFL (yellow) and Buy&Hold (blue) over the 245 test intervals in Fig. 8; also plotted in Fig. 11 is the Hang Seng Index returns (black) over the 245 test intervals as a baseline for comparison. More specifically, with $ar(i,j)$ defined in (16), the curves in Fig. 11 are

$$port(j) = \frac{1}{\sum_{i=1}^{20} w(i)} \sum_{i=1}^{20} w(i) ar(i,j) \qquad (19)$$

where $w(i)$ is the weight of the $i$'th stock in Table 1 in HSI.

To obtain a summary of the overall performance of the trading strategies for the portfolio, we take the average of each curve in Fig. 11 to get the *average annual return on portfolio* defined as:

$$\overline{port} = \frac{1}{245} \sum_{j=1}^{245} port(j) \qquad (20)$$

and its standard deviation:

$$sdv(port) = \left(\frac{1}{245}\sum_{j=1}^{245}(port(j) - \overline{port})^2\right)^{\frac{1}{2}} \quad (21)$$

Table 2 gives the $\overline{port}$, $sdv(port)$ and Sharpe ration $\overline{port}/sdv(port)$ of FollowBB, RideMood, TrendFL and Buy&Hold, and the corresponding values of HSI.

From Table 2 we see that the HSI return (index fund) is the worst among the five strategies, with an average annual return of 3.26% and Sharpe ratio 0.24, indicating that the overall market conditions in Hong Kong during this period (03-July-2007 to 02-July-2014) were very poor. The next worst performer is Buy&Hold, with an average annual return of 4.84% and Sharpe ratio 0.28, followed by TrendFL with an average annual return of 5.57% and Sharpe ratio 0.45. The best performer is RideMood, with average annual return 10.68% and Sharpe ratio 0.67, followed by FollowBB with average annual return 8.12% and Sharpe ratio 0.53. These numbers show that the portfolios with our FollowBB and RideMood strategies are two to three times better than the whole market (HSI) --- FollowBB return / HSI return = 8.12% / 3.26% ≈ 2.49, FollowBB Sharpe ratio / HSI Sharpe ratio = 0.53 / 0.24 ≈ 2.21, RideMood return / HSI return = 10.68% / 3.26% ≈ 3.27, and RideMood Sharpe ratio / HSI Sharpe ratio = 0.67 / 0.24 ≈ 2.79. Furthermore, if a fund manager switched from Buy&Hold to FollowBB or RideMood, then on average the profits would increase (8.12% - 4.84%)/4.84% ≈ 67% or (10.68% - 4.84%)/4.84% ≈ 120%, respectively, and the risk, measured by the standard deviation, would decrease |15.11% - 17.05%|/17.05% ≈ 11% or |15.43% - 17.05%|/17.05% ≈ 9.5%, respectively --- doubling the profits with less risk.

Up to now the test intervals were all two years. We now show the daily market values of the portfolios for the whole seven years from 03-July-2007 to 02-July-2014. Consider the portfolio scheme in Fig. 10 and let $V(i,t)$ be the market value of the investment on stock $i$ at trading day $t$ after deducting the transaction costs. Suppose there have been $N_{i,t}$ completed buy-sell cycles before day $t$ for stock $i$ and let $p_{i,k}^{buy}$ ($p_{i,k}^{sell}$) be the buy (sell) price of the $k$'th buy-sell cycle for stock $i$. Then, if the investment on stock $i$ at day $t$ is in the "cash" status in Figs. 5 to 7, we have

$$V(i,t) = \frac{IM \times w(i)}{\sum_{j=1}^{20} w(j)} \left(\prod_{k=1}^{N_{i,t}} \frac{p_{i,k}^{sell}}{p_{i,k}^{buy}}\right)(1 - 0.108\%)^{N_{i,t}}$$
$$\times (1 - 0.208\%)^{N_{i,t}} \quad (22)$$

where $\frac{IM \times w(i)}{\sum_{j=1}^{20} w(j)}$ is the initial money given to stock $i$ and 0.108% (0.208%) is the buy-side (sell-side) cost (see Footnote 6); if the investment on stock $i$ at day $t$ is in the "stock" status in Figs. 5 to 7, we have

$$V(i,t) = \frac{IM \times w(i)}{\sum_{j=1}^{20} w(j)} \left(\prod_{k=1}^{N_{i,t}} \frac{p_{i,k}^{sell}}{p_{i,k}^{buy}}\right)\left(\frac{p_{i,t}}{p_{i,N_{i,t}+1}^{buy}}\right)$$
$$\times (1 - 0.108\%)^{N_{i,t}+1}(1 - 0.208\%)^{N_{i,t}} \quad (23)$$

where $p_{i,t}$ is the price of stock $i$ at day $t$. According to our simple portfolio scheme in Fig. 10, the market value of the portfolio at day $t$, $V(port,t)$, is simply the summation of the $V(i,t)$'s of the 20 stocks:

$$V(port, t) = \sum_{i=1}^{20} V(i,t) \quad (24)$$

Table 2: Portfolio performance of FollowBB, RideMood, TrendFL and Buy&Hold, and the average annual Hang Seng Index (HSI) return over the 245 test intervals in Fig. 8. The returns are net returns after deducting the transaction costs.

|  | Average Annual Return of Portfolio (%) ±Standard deviation (%); Sharpe ratio |
|---|---|
| **FollowBB** | 8.1276 ±15.1144; 0.5377 |
| **RideMood** | 10.4874 ±15.4337; 0.6795 |
| **TrendFL** | 5.5739 ±12.2703; 0.4543 |
| **Buy&Hold** | 4.8429 ±17.0508; 0.2840 |
| **HSI** | 3.2691 ±13.3483; 0.2449 |

Fig. 12 plots the $V(port,t)$ for FollowBB (green), RideMood (red), TrendFL (yellow), and Buy&Hold (blue), and the HSI (black) from 03-July-2007 to 02-July-2014 with initial money IM=100. From Fig. 12 we see that RideMood has the best overall performance, followed by FollowBB, and both of which are much better than the other three, with Buy&Hold and HSI the worst performers.

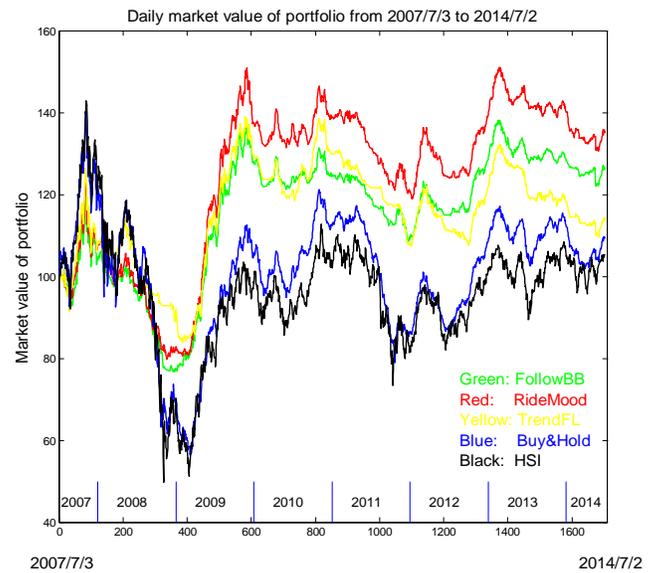

Fig. 12: Daily market value of the portfolios (after dudecting the transaction costs) using FollowBB (green), RideMood (red), TrendFL (Yellow) and Buy&Hold (blue), and HSI (black) from 03-07-2007 to 02-07-2014.

## VII. DETAILS OF BUY/SELL CYCLES

To see the buy/sell details of our FollowBB and RideMood strategies, we show, in Figs. 13.1 to 13.10 and Tables 3.1 and 3.2 for FollowBB, and in Figs. 14.1 to 14.10 and Tables 4.1 and 4.2 for RideMood, the buy/sell details of the 10 top stocks in Table 1 over the three years from 02-Jan-2011 to 31-Dec-2013, where the top sub-figures in Figs. 13.1 to 14.10 plot the daily closing prices of the 10 stocks (each figure for one stock) over the test interval and the buy (vertical green lines) sell (vertical red lines) points, the bottom sub-figures in Figs. 13.1 to 13.10 plot the 3-day moving averages $-\bar{a}_6(t,3)$ and $\bar{a}_7(t,3)$, the bottom sub-figures in Figs. 14.1 to 14.10 plot the 5-day moving averages $\overline{mood}(t,5) = \bar{a}_7(t,5) - \bar{a}_6(t,5)$, Tables 3.1 and 3.2 give the details of the buy/sell dates, prices and returns for the buy/sell cycles using FollowBB, and Tables 4.1 and 4.2 gives the returns of the buy/sell cycles using RideMood. More specifically, the green lines in the bottom sub-figures of Figs. 13.1 to 13.10 are the positive parts of $\bar{a}_7(t,3)$, indicating the presence of big buyers; the red lines in these sub-figures are the negative parts of $-\bar{a}_6(t,3)$, indicating the presence of big sellers; and the yellow lines in the figures are either the $\bar{a}_7(t,3)$ in negative or the $-\bar{a}_6(t,3)$ in positive, indicating the absence of big buyer or big seller and the dominance of trend followers. Similarly, the green (red) lines in the bottom sub-figures of Figs. 14.1 to 14.10 are the positive (negative) parts of $\overline{mood}(t,5) = \bar{a}_7(t,5) - \bar{a}_6(t,5)$, indicating that the big buyer (seller) strength is larger than the big seller (buyer) strength.

Aligning the top and bottom sub-figures in each of Figs. 13.1 to 14.10 we can see in what scenarios the buy or sell actions are committed (Tables 3.1 to 4.2 give the details of these buy-sell cycles), which give us useful information about the strengths (catching up the price increases or avoiding the price declines) and the weakness (negative return buy/sell cycles or missing the price rise opportunities) of the FollowBB and RideMood strategies. More specifically, we see from these figures and tables that our FollowBB and RideMood strategies can in general:
- Detect the major up-trends and follow them up to make big profits; and,
- Get out of the major down-trends to avoid big losses.

Generally speaking, the losses are many but small, whereas the gains are few but big, which agrees with the general experience of the successful investors [8], [18], [24], [30], [32], [36], [42].

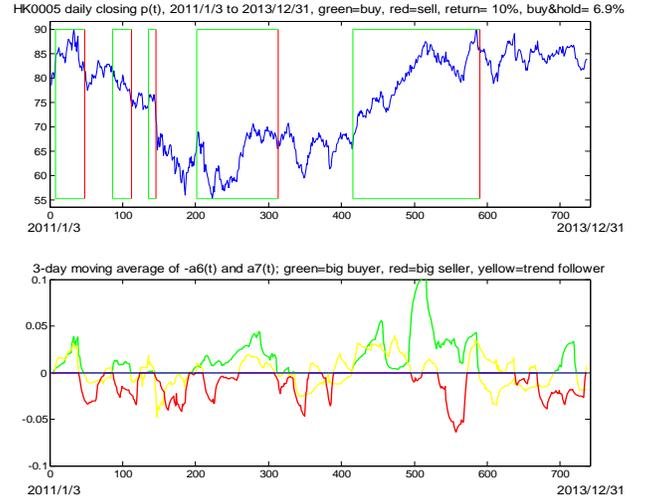

Fig. 13.1: Top: HK0005 daily closing $p_t$ from 2011-01-03 to 2013-12-31 and buy (green), sell (red) points using FollowBB. Bottom: 3-day moving average of $-a_6(t)$ and $a_7(t)$; green=big buyer ($\bar{a}_7(t,3) > 0$); red=big seller ($-\bar{a}_6(t,3) < 0$); yellow=trend follower ($\bar{a}_7(t,3) < 0$ or $-\bar{a}_6(t,3) > 0$).

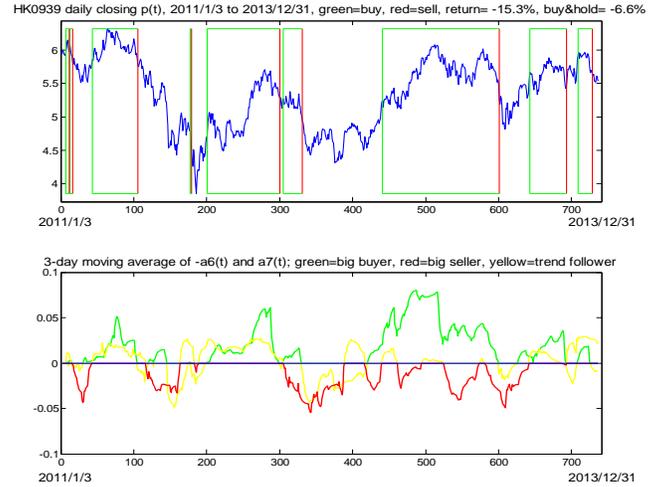

Fig. 13.2: Same as Fig. 13.1 for HK0939 (FollowBB).

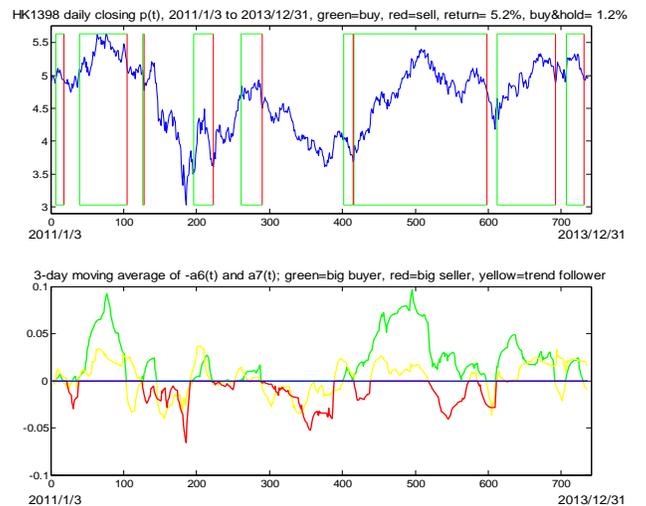

Fig. 13.3: Same as Fig. 13.1 for HK1398 (FollowBB).

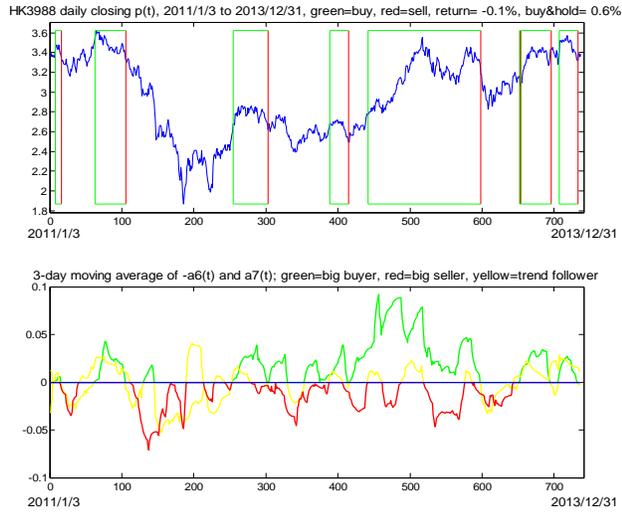

Fig. 13.4: Same as Fig. 13.1 for HK3988 (FollowBB).

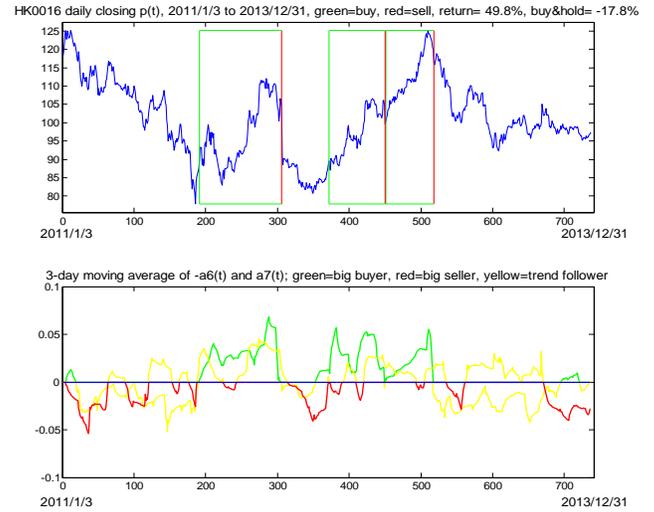

Fig. 13.7: Same as Fig. 13.1 for HK0016 (FollowBB).

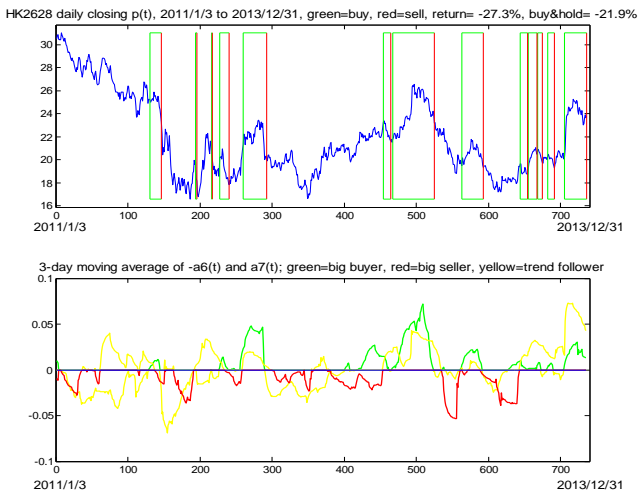

Fig. 13.5: Same as Fig. 13.1 for HK2628 (FollowBB).

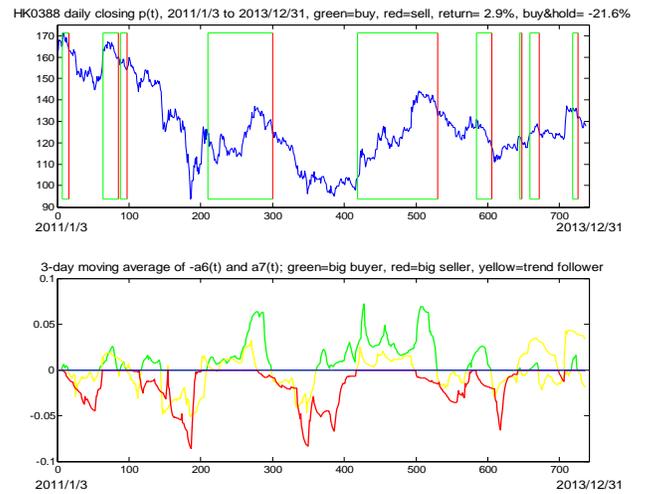

Fig. 13.8: Same as Fig. 13.1 for HK0388 (FollowBB).

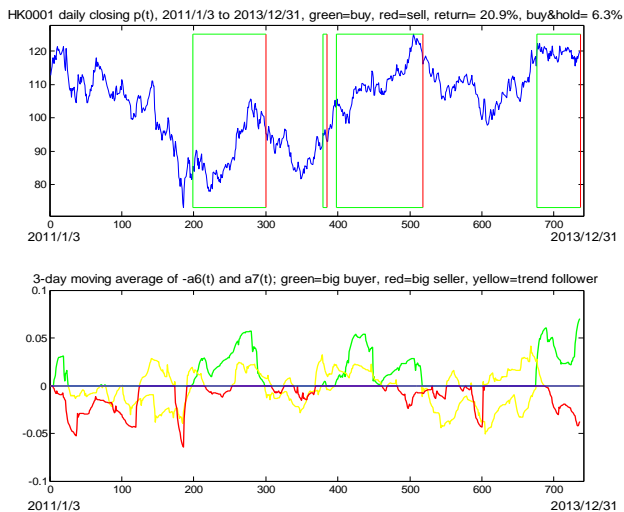

Fig. 13.6: Same as Fig. 13.1 for HK0001 (FollowBB).

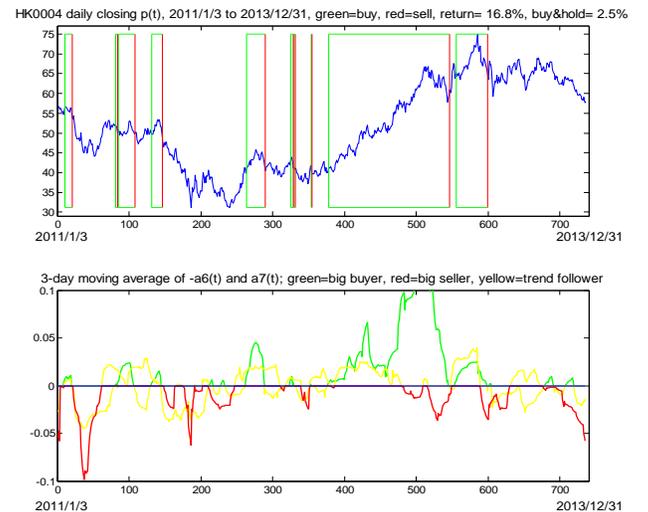

Fig. 13.9: Same as Fig. 13.1 for HK0004 (FollowBB).

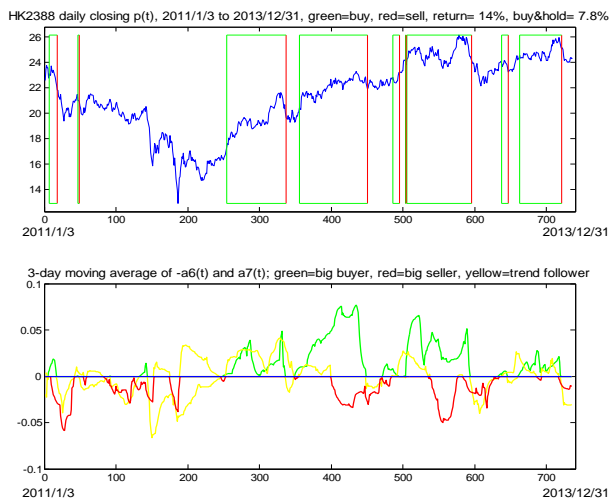

Fig. 13.10: Same as Fig. 13.1 for HK2388 (FollowBB).

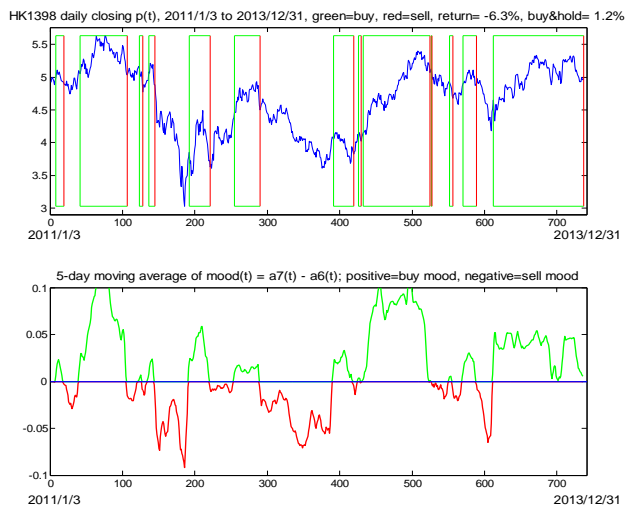

Fig. 14.3: Same as Fig. 14.1 for HK1398 (RideMood).

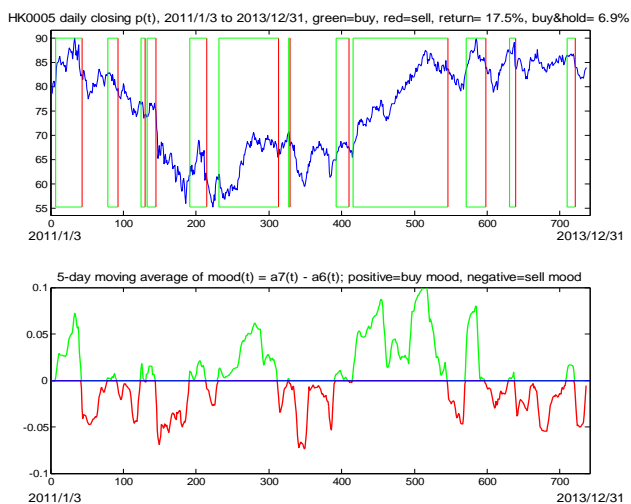

Fig. 14.1: Top: HK0005 daily closing $p_t$ 2011/1/3 to 2013/12/31 and buy (green) sell (red) points using RideMood. Bottom: 5-day moving average of $a_7(t) - a_6(t)$; green: $\bar{a}_7(t,5) > \bar{a}_6(t,5)$ (big buyer stronger than big seller); red: $\bar{a}_7(t,5) < \bar{a}_6(t,5)$ (big buyer weaker than big seller).

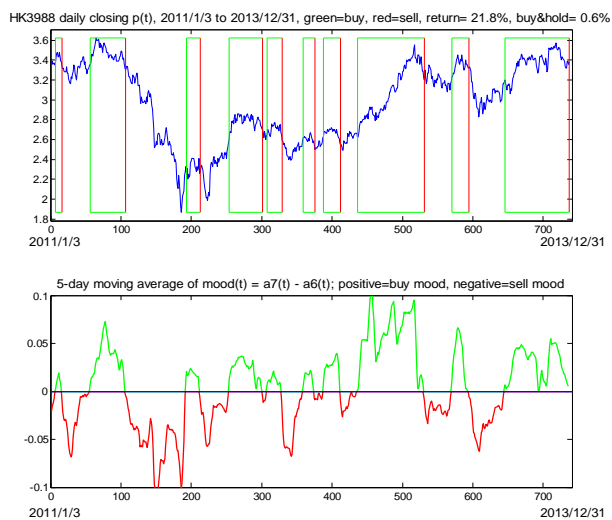

Fig. 14.4: Same as Fig. 14.1 for HK3988 (RideMood).

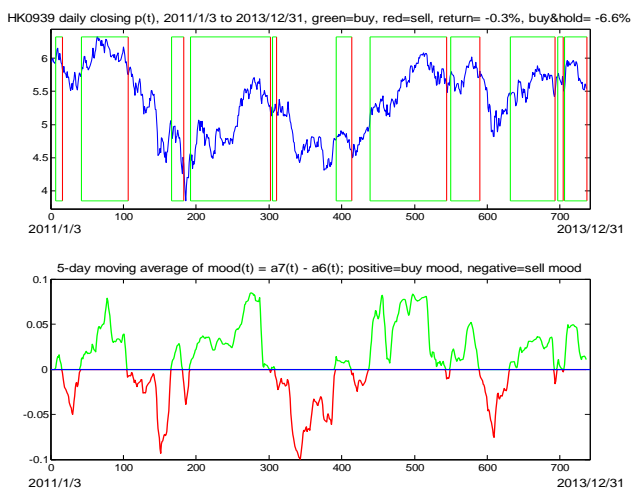

Fig. 14.2: Same as Fig. 14.1 for HK0939 (RideMood).

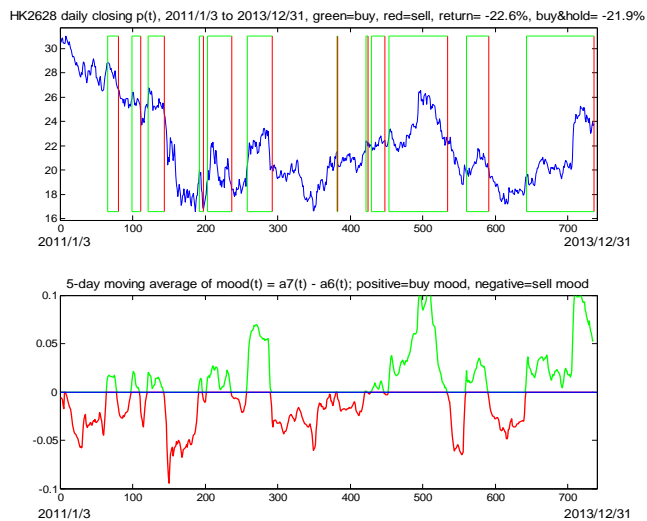

Fig. 14.5: Same as Fig. 14.1 for HK2628 (RideMood).

**Table 3.1: Details of buy-sell cycles using FollowBB over the three years from 2011-01-03 to 2013-12-31 for HK0005, HK0939, HK1398, HK3988 and HK2628 (corresponding to Figs. 13.1 to 13.5). The returns are net returns after deducting transaction costs.**

| Cycle No. | HK0005 | HK0939 | HK1398 | HK3988 | HK2628 |
|---|---|---|---|---|---|
| 1 | Buy: 83.81; 2011/1/12<br>Sell: 81.26; 2011/3/11<br>Return: -3.0426% | Buy: 5.97; 2011/1/11<br>Sell: 5.98; 2011/1/18<br>Return: 0.1675% | Buy: 4.95; 2011/1/11<br>Sell: 4.93; 2011/1/26<br>Return: -0.40404% | Buy: 3.43; 2011/1/12<br>Sell: 3.33; 2011/1/24<br>Return: -2.9155% | Buy: 25.18; 2011/7/15<br>Sell: 23.26; 2011/8/5<br>Return: -7.6251% |
| 2 | Buy: 81.38; 2011/5/11<br>Sell: 74.79; 2011/6/17<br>Return: -8.0978% | Buy: 6.07; 2011/1/19<br>Sell: 5.82; 2011/1/24<br>Return: -4.1186% | Buy: 5.1; 2011/3/1<br>Sell: 5.11; 2011/6/8<br>Return: 0.19608% | Buy: 3.54; 2011/4/1<br>Sell: 3.32; 2011/6/9<br>Return: -6.2147% | Buy: 19.79; 2011/10/17<br>Sell: 18.16; 2011/10/19<br>Return: -8.2365% |
| 3 | Buy: 74.64; 2011/7/21<br>Sell: 71.49; 2011/8/5<br>Return: -4.2203% | Buy: 5.86; 2011/3/4<br>Sell: 5.77; 2011/6/9<br>Return: -1.5358% | Buy: 5.08; 2011/7/8<br>Sell: 4.77; 2011/7/12<br>Return: -6.1024% | Buy: 2.68; 2012/1/13<br>Sell: 2.62; 2012/3/26<br>Return: -2.2388% | Buy: 20.46; 2011/11/16<br>Sell: 20.03; 2011/11/17<br>Return: -2.1017% |
| 4 | Buy: 66.27; 2011/10/27<br>Sell: 65.72; 2012/4/12<br>Return: -0.82994% | Buy: 4.8; 2011/9/20<br>Sell: 4.45; 2011/9/22<br>Return: -7.2917% | Buy: 3.57; 2011/10/19<br>Sell: 3.61; 2011/11/25<br>Return: 1.1204% | Buy: 2.68; 2012/7/31<br>Sell: 2.49; 2012/9/5<br>Return: -7.0896% | Buy: 20.22; 2011/12/1<br>Sell: 17.81; 2011/12/20<br>Return: -11.9189% |
| 5 | Buy: 66.63; 2012/9/6<br>Sell: 86.58; 2013/5/28<br>Return: 29.9415% | Buy: 4.71; 2011/10/26<br>Sell: 5.27; 2012/3/21<br>Return: 11.8896% | Buy: 4.8; 2012/1/26<br>Sell: 4.5; 2012/3/7<br>Return: -6.25% | Buy: 2.77; 2012/10/16<br>Sell: 3.18; 2013/6/7<br>Return: 14.8014% | Buy: 21.33; 2012/1/20<br>Sell: 20.17; 2012/3/9<br>Return: -5.4383% |
| 6 | | Buy: 5.27; 2012/3/28<br>Sell: 4.94; 2012/5/9<br>Return: -6.2619% | Buy: 4.06; 2012/8/17<br>Sell: 3.68; 2012/9/5<br>Return: -9.3596% | Buy: 3.14; 2013/8/27<br>Sell: 3.06; 2013/8/28<br>Return: -2.5478% | Buy: 23.26; 2012/11/2<br>Sell: 21.6; 2012/11/16<br>Return: -7.1367% |
| 7 | | Buy: 5.25; 2012/10/15<br>Sell: 5.25; 2013/6/13<br>Return: 0% | Buy: 3.71; 2012/9/6<br>Sell: 4.74; 2013/6/7<br>Return: 27.7628% | Buy: 3.11; 2013/8/29<br>Sell: 3.44; 2013/10/31<br>Return: 10.6109% | Buy: 21.8; 2012/11/21<br>Sell: 23.16; 2013/2/19<br>Return: 6.2385% |
| 8 | | Buy: 5.67; 2013/8/13<br>Sell: 5.47; 2013/10/28<br>Return: -3.5273% | Buy: 4.65; 2013/6/28<br>Sell: 4.88; 2013/10/25<br>Return: 4.9462% | Buy: 3.4; 2013/11/15<br>Sell: 3.32; 2013/12/23<br>Return: -2.3529% | Buy: 19.51; 2013/4/18<br>Sell: 19.35; 2013/6/3<br>Return: -0.82009% |
| 9 | | Buy: 5.92; 2013/11/19<br>Sell: 5.65; 2013/12/16<br>Return: -4.5608% | Buy: 5.06; 2013/11/15<br>Sell: 4.93; 2013/12/19<br>Return: -2.5692% | | Buy: 19.63; 2013/8/16<br>Sell: 18.7; 2013/8/30<br>Return: -4.7376% |
| 10 | | | | | Buy: 19.19; 2013/9/2<br>Sell: 20.48; 2013/9/18<br>Return: 6.7223% |
| 11 | | | | | Buy: 20.82; 2013/9/19<br>Sell: 19.74; 2013/9/30<br>Return: -5.1873% |
| 12 | | | | | Buy: 20.33; 2013/10/11<br>Sell: 19.27; 2013/10/25<br>Return: -5.214% |
| 13 | | | | | Buy: 20.24; 2013/11/14<br>Sell: 23.82; 2013/12/31<br>Return: 17.6877% |
| Accumulated Return | **10%** | **-15.3%** | **5.2%** | **-0.1%** | **-27.1%** |
| Buy&Hold Return | **6.9%** | **-6.6%** | **1.2%** | **0.6%** | **-21.9%** |

                                                                                                                                                                          15

**Table 3.2: Details of buy-sell cycles using FollowBB over the three years from 2011-01-03 to 2013-12-31 for HK0001, HK0016, HK0388, HK0004 and HK2388 (corresponding to Figs. 13.6 -- 13.10). The returns are net returns after deducting transaction costs.**

| Cycle No. | HK0001 | HK0016 | HK0388 | HK0004 | HK2388 |
|---|---|---|---|---|---|
| 1 | Buy: 85.42; 2011/10/24<br>Sell: 95.27; 2012/3/21<br>Return: 11.5313% | Buy: 89.57; 2011/10/12<br>Sell: 104.24; 2012/3/29<br>Return: 16.3783% | Buy: 165.57; 2011/1/11<br>Sell: 162.67; 2011/1/24<br>Return: -1.7515% | Buy: 56.55; 2011/1/17<br>Sell: 53.81; 2011/1/31<br>Return: -4.8453% | Buy: 23.11; 2011/1/11<br>Sell: 21.84; 2011/1/26<br>Return: -5.4955% |
| 2 | Buy: 95.44; 2012/7/17<br>Sell: 92.89; 2012/7/25<br>Return: -2.6718% | Buy: 89.65; 2012/7/6<br>Sell: 99.55; 2012/10/29<br>Return: 11.0429% | Buy: 163.85; 2011/4/4<br>Sell: 160.28; 2011/5/9<br>Return: -2.1788% | Buy: 52.08; 2011/5/3<br>Sell: 49.75; 2011/5/6<br>Return: -4.4739% | Buy: 21; 2011/3/10<br>Sell: 20.91; 2011/3/14<br>Return: -0.42857% |
| 3 | Buy: 104.06; 2012/8/13<br>Sell: 116.08; 2013/2/5<br>Return: 11.551% | Buy: 100.3; 2012/10/30<br>Sell: 116.25; 2013/2/5<br>Return: 15.9023% | Buy: 158.81; 2011/5/13<br>Sell: 154.68; 2011/5/26<br>Return: -2.6006% | Buy: 50.8; 2011/5/9<br>Sell: 50; 2011/6/13<br>Return: -1.5748% | Buy: 17.37; 2012/1/12<br>Sell: 19.7; 2012/5/17<br>Return: 13.4139% |
| 4 | Buy: 119.98; 2013/10/2<br>Sell: 119.78; 2013/12/31<br>Return: -0.16669% | | Buy: 123.18; 2011/11/8<br>Sell: 124.21; 2012/3/21<br>Return: 0.83617% | Buy: 49.67; 2011/7/18<br>Sell: 48.79; 2011/8/5<br>Return: -1.7717% | Buy: 20.43; 2012/6/13<br>Sell: 21.92; 2012/10/29<br>Return: 7.2932% |
| 5 | | | Buy: 103.69; 2012/9/10<br>Sell: 132.67; 2013/2/26<br>Return: 27.9487% | Buy: 41.09; 2012/1/31<br>Sell: 40.48; 2012/3/7<br>Return: -1.4845% | Buy: 22.81; 2012/12/19<br>Sell: 23.23; 2013/1/4<br>Return: 1.8413% |
| 6 | | | Buy: 130.94; 2013/5/20<br>Sell: 117.89; 2013/6/20<br>Return: -9.9664% | Buy: 43.14; 2012/4/30<br>Sell: 41.04; 2012/5/7<br>Return: -4.8679% | Buy: 24.12; 2013/1/16<br>Sell: 23.84; 2013/1/17<br>Return: -1.1609% |
| 7 | | | Buy: 122.86; 2013/8/16<br>Sell: 120.33; 2013/8/20<br>Return: -2.0593% | Buy: 41.37; 2012/5/8<br>Sell: 40.76; 2012/5/9<br>Return: -1.4745% | Buy: 24.26; 2013/1/18<br>Sell: 24.17; 2013/6/5<br>Return: -0.37098% |
| 8 | | | Buy: 124.79; 2013/9/5<br>Sell: 124.89; 2013/9/25<br>Return: 0.080135% | Buy: 40.56; 2012/6/12<br>Sell: 39.75; 2012/6/13<br>Return: -1.997% | Buy: 23.89; 2013/8/6<br>Sell: 23.26; 2013/8/21<br>Return: -2.6371% |
| 9 | | | Buy: 136.16; 2013/12/2<br>Sell: 131.41; 2013/12/12<br>Return: -3.4885% | Buy: 41.18; 2012/7/17<br>Sell: 60.9; 2013/3/21<br>Return: 47.8873% | Buy: 24.56; 2013/9/11<br>Sell: 25.04; 2013/12/6<br>Return: 1.9544% |
| 10 | | | | Buy: 65.15; 2013/4/8<br>Sell: 64.7; 2013/6/11<br>Return: -0.69071% | |
| Accumulated Return | **20.9%** | **49.8%** | **2.9%** | **16.8%** | **14%** |
| Buy&Hold Return | **6.3%** | **-17.8%** | **-21.6%** | **2.5%** | **7.8%** |

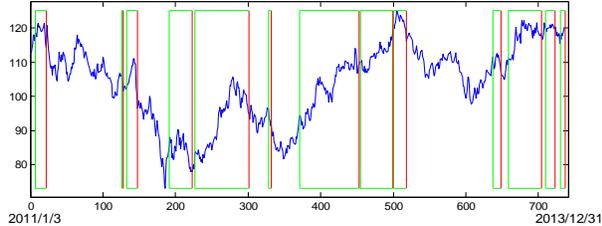

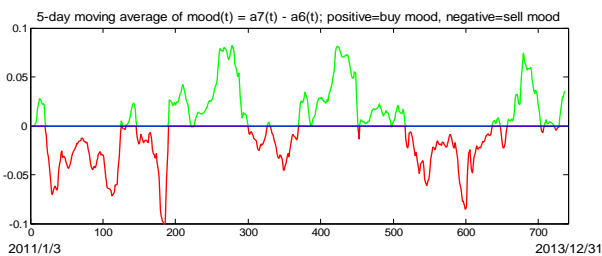

Fig. 14.6: Same as Fig. 14.1 for HK0001 (RideMood).

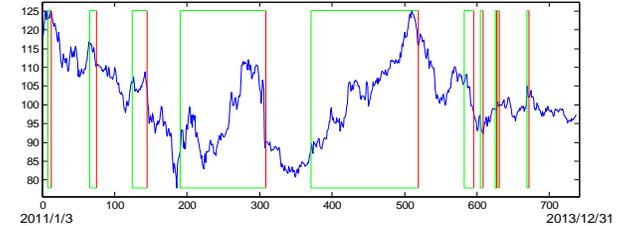

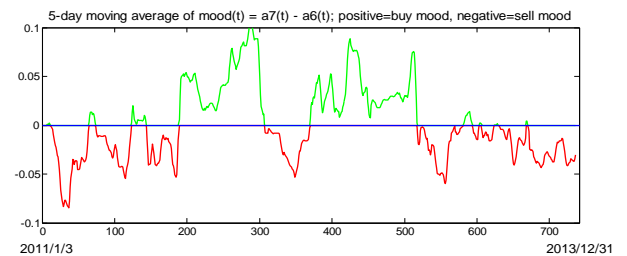

Fig. 14.7: Same as Fig. 14.1 for HK0016 (RideMood).

                              16

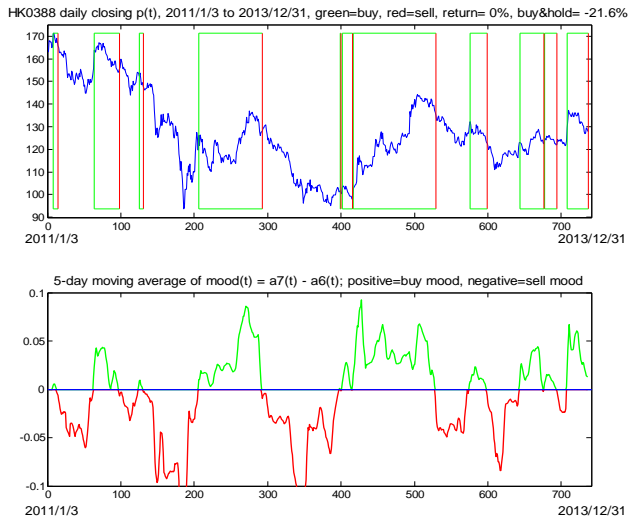

Fig. 14.8: Same as Fig. 14.1 for HK0388 (RideMood).

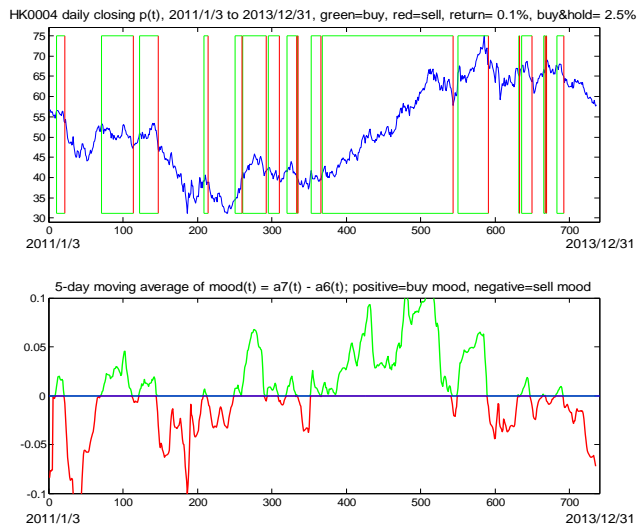

Fig. 14.9: Same as Fig. 14.1 for HK0004 (RideMood).

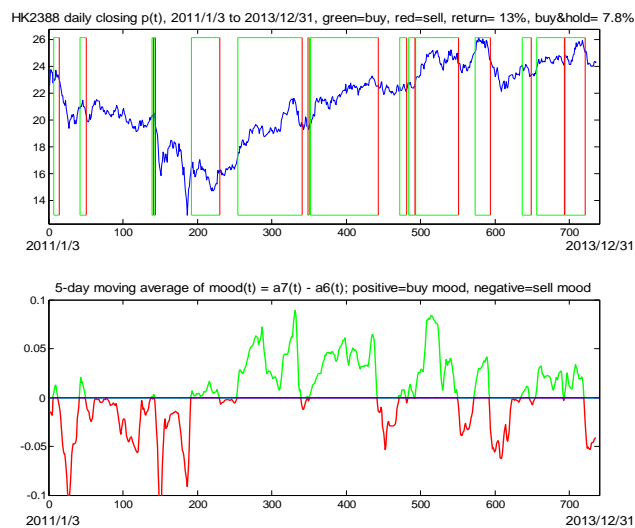

Fig. 14.10: Same as Fig. 14.1 for HK2388 (RideMood).

## VIII. CONCLUDING REMARKS

Based on the testing for the top 20 banking and real estate stocks listed in the Hong Kong Stock Exchange over the 245 two-year test intervals uniformly distributed across the seven years from 03-July-2007 to 02-July-2014 which includes a variety of scenarios such as the strong rise in late 2007 due to "irrational exuberance" [40], the big decline during the 2008 financial crisis [29], the recoverary in 2009 after the panic [7] and the "ordinary" [41] years from 2010 to 2014, the average annual returns (±Standard deviation) of the Follow-the-Big-Buyer (FollowBB), Ride-the-Mood (RideMood) and Buy-and-Hold (Buy&Hold) strategies, after deducting the transaction costs, are 8.12% (±15.11%), 10.48% (±15.43%) and 4.84% (±17.05%), respectively. This means that if a fund manager switched from the Buy&Hold strategy to the FollowBB or RideMood strategies during this period, the net profits on average would increase (8.12% - 4.84%)/4.84% ≈ 67% or (10.68% - 4.84%)/4.84% ≈ 120%, respectively, and the risk, measured by the standard deviation, would decrease |15.11% - 17.05%|/17.05% ≈ 11% or |15.43% - 17.05%|/17.05% ≈ 9.5%, respectively. For funds with large sums of money, these profit advances and risk reductions are significant.

The FollowBB and RideMood strategies proposed in this paper have the following characteristics:

- They are very general. The only information they need is the price data of an asset up to the current time: $\{p_0, p_1, ..., p_t\}$. The asset can be anything: the stocks of different countries, commodities, foreign exchange products, derivatives, …; and the sampling interval can be one-day, one-hour, 10-minute, 1-minute, 3-second, …. As long as the traders of the asset use technical analysis in their decision making processes, the models of this paper are relevant and the trading strategies could be tried.

- They are easy to implement. You simply put the price data $\{p_0, p_1, ..., p_t\}$ into the parameter estimation algorithm (11)-(13) and make your buy/sell decisions based on the estimated parameters according to the simple flow-charts in Figs. 5 and 6; it takes just a few lines of MATLAB codes to implement these computations.

- They have good reason to succeed. They do not predict returns [38], in fact they do not predict anything about the future;[7] they simply follow the footprints of the big buyers and the big sellers: if the big buyers succeed, they succeed; if the big buyers fail, they fail. As long as you believe supply-demand determines prices [6], their philosophy is well-justified.

---

[7] "It is hard to make predictions, especially about the future." --- Mark Twain.

For referencing please quote: *IEEE Trans. on Fuzzy Systems* 23(5): 1680-1697, 2015.

Table 4.1: Details of buy-sell cycles using RideMood over the three years from 2011-01-03 to 2013-12-31 for HK0005, HK0939, HK1398, HK3988 and HK2628 (corresponding to Figs. 14.1 to 14.5). The returns are net returns after deducting transaction costs.

| Cycle No. | HK0005 | HK0939 | HK1398 | HK3988 | HK2628 |
|---|---|---|---|---|---|
| 1 | Buy: 80.82; 2011/1/11<br>Sell: 82.29; 2011/3/4<br>Return: 1.8189% | Buy: 5.97; 2011/1/11<br>Sell: 5.82; 2011/1/24<br>Return: -2.5126% | Buy: 4.99; 2011/1/12<br>Sell: 4.9; 2011/1/27<br>Return: -1.8036% | Buy: 3.37; 2011/1/11<br>Sell: 3.33; 2011/1/24<br>Return: -1.1869% | Buy: 28.86; 2011/4/7<br>Sell: 26.58; 2011/4/29<br>Return: -7.9002% |
| 2 | Buy: 82.96; 2011/4/28<br>Sell: 79.66; 2011/5/20<br>Return: -3.9778% | Buy: 5.79; 2011/3/3<br>Sell: 5.75; 2011/6/10<br>Return: -0.69085% | Buy: 5.05; 2011/3/2<br>Sell: 5.03; 2011/6/10<br>Return: -0.39604% | Buy: 3.35; 2011/3/23<br>Sell: 3.3; 2011/6/10<br>Return: -1.4925% | Buy: 25.9; 2011/5/30<br>Sell: 24.7; 2011/6/16<br>Return: -4.6332% |
| 3 | Buy: 76.46; 2011/7/6<br>Sell: 74.25; 2011/7/14<br>Return: -2.8904% | Buy: 4.98; 2011/9/2<br>Sell: 4.4; 2011/9/28<br>Return: -11.6466% | Buy: 5.14; 2011/7/5<br>Sell: 4.77; 2011/7/12<br>Return: -7.1984% | Buy: 2.26; 2011/10/14<br>Sell: 2.28; 2011/11/11<br>Return: 0.88496% | Buy: 25.61; 2011/6/30<br>Sell: 24.6; 2011/8/3<br>Return: -3.9438% |
| 4 | Buy: 74; 2011/7/19<br>Sell: 74.89; 2011/8/4<br>Return: 1.2027% | Buy: 4.57; 2011/10/13<br>Sell: 5.19; 2012/3/23<br>Return: 13.5667% | Buy: 5.08; 2011/7/22<br>Sell: 4.88; 2011/8/4<br>Return: -3.937% | Buy: 2.6; 2012/1/12<br>Sell: 2.7; 2012/3/22<br>Return: 3.8462% | Buy: 19.98; 2011/10/13<br>Sell: 16.77; 2011/10/21<br>Return: -16.0661% |
| 5 | Buy: 61.64; 2011/10/12<br>Sell: 61.1; 2011/11/15<br>Return: -0.87605% | Buy: 5.27; 2012/3/28<br>Sell: 5.2; 2012/4/5<br>Return: -1.3283% | Buy: 3.85; 2011/10/13<br>Sell: 3.63; 2011/11/23<br>Return: -5.7143% | Buy: 2.65; 2012/4/2<br>Sell: 2.59; 2012/5/7<br>Return: -2.2642% | Buy: 19.24; 2011/10/28<br>Sell: 18.33; 2011/12/15<br>Return: -4.7297% |
| 6 | Buy: 62.03; 2011/12/7<br>Sell: 65.72; 2012/4/12<br>Return: 5.9487% | Buy: 4.78; 2012/8/3<br>Sell: 4.58; 2012/9/4<br>Return: -4.1841% | Buy: 4.46; 2012/1/13<br>Sell: 4.5; 2012/3/7<br>Return: 0.89686% | Buy: 2.62; 2012/6/18<br>Sell: 2.5; 2012/7/12<br>Return: -4.5802% | Buy: 20.32; 2012/1/18<br>Sell: 20.17; 2012/3/9<br>Return: -0.73819% |
| 7 | Buy: 70.81; 2012/5/3<br>Sell: 67.94; 2012/5/7<br>Return: -4.0531% | Buy: 5.13; 2012/10/11<br>Sell: 5.56; 2013/3/18<br>Return: 8.3821% | Buy: 4.06; 2012/8/2<br>Sell: 3.82; 2012/9/11<br>Return: -5.9113% | Buy: 2.63; 2012/7/30<br>Sell: 2.54; 2012/8/31<br>Return: -3.4221% | Buy: 21.51; 2012/7/20<br>Sell: 20.48; 2012/7/23<br>Return: -4.7885% |
| 8 | Buy: 64.91; 2012/8/3<br>Sell: 67.57; 2012/8/29<br>Return: 4.098% | Buy: 5.77; 2013/3/26<br>Sell: 5.81; 2013/5/28<br>Return: 0.69324% | Buy: 4.08; 2012/9/20<br>Sell: 4.02; 2012/9/26<br>Return: -1.4706% | Buy: 2.62; 2012/10/8<br>Sell: 3.2; 2013/2/27<br>Return: 22.1374% | Buy: 22.38; 2012/9/14<br>Sell: 21.99; 2012/9/18<br>Return: -1.7426% |
| 9 | Buy: 66.63; 2012/9/6<br>Sell: 83.55; 2013/3/20<br>Return: 25.394% | Buy: 5.43; 2013/7/29<br>Sell: 5.47; 2013/10/28<br>Return: 0.73665% | Buy: 4.15; 2012/9/28<br>Sell: 5.07; 2013/2/19<br>Return: 22.1687% | Buy: 3.23; 2013/4/26<br>Sell: 3.32; 2013/6/3<br>Return: 2.7864% | Buy: 22.09; 2012/9/25<br>Sell: 22.43; 2012/10/25<br>Return: 1.5392% |
| 10 | Buy: 83.55; 2013/4/29<br>Sell: 83.79; 2013/6/7<br>Return: 0.28725% | Buy: 5.74; 2013/11/1<br>Sell: 5.66; 2013/11/12<br>Return: -1.3937% | Buy: 5.13; 2013/2/20<br>Sell: 5.02; 2013/2/21<br>Return: -2.1442% | Buy: 3.2; 2013/8/19<br>Sell: 3.38; 2013/12/31<br>Return: 5.625% | Buy: 22.82; 2012/11/1<br>Sell: 22.14; 2013/3/4<br>Return: -2.9798% |
| 11 | Buy: 87.68; 2013/7/26<br>Sell: 83.69; 2013/8/7<br>Return: -4.5506% | Buy: 5.54; 2013/11/14<br>Sell: 5.55; 2013/12/31<br>Return: 0.18051% | Buy: 4.92; 2013/3/28<br>Sell: 4.68; 2013/4/8<br>Return: -4.878% | | Buy: 20.82; 2013/9/19<br>Sell: 19.74; 2013/9/30<br>Return: -5.1873 |
| 12 | Buy: 86.49; 2013/11/20<br>Sell: 83.99; 2013/12/5<br>Return: -2.8905% | | Buy: 4.9; 2013/4/26<br>Sell: 4.9; 2013/5/27<br>Return: 0% | | Buy: 19.41; 2013/8/15<br>Sell: 23.82; 2013/12/31<br>Return: 22.7202% |
| 13 | | | Buy: 4.65; 2013/6/28<br>Sell: 4.98; 2013/12/31<br>Return: 7.0968% | | |
| Accumulated Return | **17.5%** | **-0.3%** | **-6.3%** | **21.8%** | **-22.6%** |
| Buy&Hold Return | **6.9%** | **-6.6%** | **1.2%** | **0.6%** | **-21.9%** |

There is a famous joke widely told among economists [28]: One day an economics professor and his student were strolling down the street and they came upon a $100 bill lying on the ground, and as the student reached down to pick it up, the professor said: "Don't bother --- if it were a genuine $100 bill, someone would have already picked it up"[8]. Just because so many people are brainwashed by this Efficient Market Hypothesis [14], [33], there are many real $100 bills lying on the ground untouched. So go ahead and pick them up [11], [26] (or you may try the methods of this paper), they are real money [42], [44].

At the end of this three-part paper, it may be helpful to discuss some future research directions[7]. Fig. 15 illustrates the skeleton of this study from which we see at least the following research opportunities:

a. Transforming more technical trading rules into price dynamical equations. As emphasized in [45], [46], the main contribution of fuzzy systems theory is to provide a convenient mathematical framework (fuzzy sets, fuzzy IF-THEN rules, fuzzy inference engines, etc.) to transform linguistic descriptions into mathematical formulas.

---

[8] The story ended something like this: The student picked up the $100 bill and whispered to the professor: "I am the 'someone' you were talking about, Sir."

Table 4.2: Details of buy-sell cycles using RideMood over the three years from 2011-01-03 to 2013-12-31 for HK0001, HK0016, HK0388, HK0004 and HK2388 (corresponding to Figs. 14.6 -- 14.10). The returns are net returns after deducting transaction costs.

| Cycle No. | HK0001 | HK0016 | HK0388 | HK0004 | HK2388 |
|---|---|---|---|---|---|
| 1 | Buy: 118.73; 2011/1/11<br>Sell: 114.25; 2011/2/1<br>Return: -3.7733% | Buy: 124.9; 2011/1/12<br>Sell: 124.63; 2011/1/19<br>Return: -0.21617% | Buy: 171.01; 2011/1/12<br>Sell: 165.85; 2011/1/20<br>Return: -3.0174% | Buy: 56.55; 2011/1/17<br>Sell: 53.63; 2011/2/1<br>Return: -5.1636% | Buy: 23.11; 2011/1/11<br>Sell: 23.02; 2011/1/20<br>Return: -0.38944% |
| 2 | Buy: 106.57; 2011/7/8<br>Sell: 101.63; 2011/7/12<br>Return: -4.6355% | Buy: 115.58; 2011/4/7<br>Sell: 110.86; 2011/4/20<br>Return: -4.0838% | Buy: 163.85; 2011/4/4<br>Sell: 155.78; 2011/5/27<br>Return: -4.9252% | Buy: 53.03; 2011/4/14<br>Sell: 47.69; 2011/6/21<br>Return: -10.0698% | Buy: 20.87; 2011/3/3<br>Sell: 20.4; 2011/3/16<br>Return: -2.252% |
| 3 | Buy: 102.54; 2011/7/19<br>Sell: 96.5; 2011/8/9<br>Return: -5.8904% | Buy: 105.68; 2011/7/6<br>Sell: 104.68; 2011/8/4<br>Return: -0.94625% | Buy: 152.39; 2011/7/7<br>Sell: 147.71; 2011/7/15<br>Return: -3.0711% | Buy: 50.97; 2011/7/4<br>Sell: 46.99; 2011/8/8<br>Return: -7.8085% | Buy: 20.33; 2011/7/27<br>Sell: 20.16; 2011/7/29<br>Return: -0.8362% |
| 4 | Buy: 85.51; 2011/10/12<br>Sell: 78.1; 2011/11/25<br>Return: -8.6657% | Buy: 87.57; 2011/10/11<br>Sell: 90.3; 2012/4/3<br>Return: 3.1175% | Buy: 126.16; 2011/11/2<br>Sell: 127.93; 2012/3/12<br>Return: 1.403% | Buy: 38.99; 2011/11/7<br>Sell: 38.76; 2011/11/14<br>Return: -0.58989% | Buy: 20.55; 2011/8/1<br>Sell: 19.55; 2011/8/3<br>Return: -4.8662% |
| 5 | Buy: 83.58; 2011/12/1<br>Sell: 95.18; 2012/3/22<br>Return: 13.8789% | Buy: 88.48; 2012/7/5<br>Sell: 115.77; 2013/2/6<br>Return: 30.8431% | Buy: 101.85; 2012/8/14<br>Sell: 100.71; 2012/8/15<br>Return: -1.1193% | Buy: 34.47; 2012/1/9<br>Sell: 39.97; 2012/1/20<br>Return: 15.9559% | Buy: 16.58; 2011/10/13<br>Sell: 16.08; 2011/12/6<br>Return: -3.0157% |
| 6 | Buy: 96.37; 2012/5/4<br>Sell: 91.77; 2012/5/10<br>Return: -4.7733% | Buy: 106.01; 2013/5/15<br>Sell: 98.09; 2013/6/4<br>Return: -7.471% | Buy: 103.08; 2012/8/17<br>Sell: 97.82; 2012/9/6<br>Return: -5.1028% | Buy: 41.32; 2012/1/26<br>Sell: 41.6; 2012/3/9<br>Return: 0.67764% | Buy: 17.37; 2012/1/12<br>Sell: 19.52; 2012/5/23<br>Return: 12.3777% |
| 7 | Buy: 94.4; 2012/7/5<br>Sell: 109.42; 2012/11/1<br>Return: 15.911% | Buy: 96.4; 2013/6/19<br>Sell: 92.29; 2013/6/24<br>Return: -4.2635% | Buy: 102.73; 2012/9/7<br>Sell: 134.99; 2013/2/25<br>Return: 31.4027% | Buy: 42.3; 2012/3/14<br>Sell: 40.02; 2012/4/5<br>Return: -5.3901% | Buy: 19.7; 2012/6/1<br>Sell: 19.38; 2012/6/5<br>Return: -1.6244% |
| 8 | Buy: 109.99; 2012/11/5<br>Sell: 118.65; 2013/1/9<br>Return: 7.8734% | Buy: 98.86; 2013/7/18<br>Sell: 98.67; 2013/7/22<br>Return: -0.19219% | Buy: 132.21; 2013/5/7<br>Sell: 122.86; 2013/6/10<br>Return: -7.0721% | Buy: 41.46; 2012/4/23<br>Sell: 39.93; 2012/5/11<br>Return: -3.6903% | Buy: 19.84; 2012/6/7<br>Sell: 22.58; 2012/10/17<br>Return: 13.8105% |
| 9 | Buy: 120.93; 2013/1/10<br>Sell: 116.08; 2013/2/5<br>Return: -4.0106% | Buy: 100.02; 2013/7/23<br>Sell: 99.93; 2013/7/26<br>Return: -0.089982% | Buy: 123.15; 2013/8/15<br>Sell: 122.42; 2013/10/2<br>Return: -0.59277% | Buy: 39.88; 2012/5/14<br>Sell: 39.51; 2012/5/15<br>Return: -0.92778% | Buy: 22.49; 2012/11/29<br>Sell: 22.3; 2012/12/12<br>Return: -0.84482% |
| 10 | Buy: 111.57; 2013/8/5<br>Sell: 107; 2013/8/21<br>Return: -4.0961% | Buy: 105.05; 2013/9/19<br>Sell: 102.05; 2013/9/25<br>Return: -2.8558% | Buy: 124.6; 2013/10/3<br>Sell: 122.23;2013/10/29<br>Return: -1.9021% | Buy: 41.7; 2012/6/11<br>Sell: 39.28; 2012/6/28<br>Return: -5.8034% | Buy: 22.77; 2012/12/17<br>Sell: 22.53; 2012/12/31<br>Return: -1.054% |
| 11 | Buy: 110.58; 2013/9/5<br>Sell: 117.34;2013/11/12<br>Return: 6.1132% | | Buy: 129.34;2013/11/18<br>Sell: 127.76;2013/12/31<br>Return: -1.2216% | Buy: 41.08; 2012/7/3<br>Sell: 57.79; 2013/3/18<br>Return: 40.6767% | Buy: 22.95; 2013/1/2<br>Sell: 24.22; 2013/3/28<br>Return: 5.5338% |
| 12 | Buy: 119.59;2013/11/20<br>Sell: 119.88; 2013/12/9<br>Return: 0.2425% | | | Buy: 65.91; 2013/3/27<br>Sell: 67.86; 2013/5/30<br>Return: 2.9586% | Buy: 25.1; 2013/5/3<br>Sell: 24.46; 2013/6/4<br>Return: -2.5498% |
| 13 | Buy: 118.02;2013/12/18<br>Sell: 119.78;2013/12/31<br>Return: 1.4913% | | | Buy: 65.29; 2013/7/30<br>Sell: 64.85; 2013/7/31<br>Return: -0.67392% | Buy: 24.08; 2013/8/5<br>Sell: 23.41; 2013/8/22<br>Return: -2.7824% |
| 14 | | | | Buy: 67.18; 2013/8/2<br>Sell: 63.68; 2013/8/23<br>Return: -5.2099% | Buy: 23.93; 2013/9/3<br>Sell: 24.41; 2013/10/29<br>Return: 2.0059% |
| 15 | | | | Buy: 66.66; 2013/9/16<br>Sell: 65.97; 2013/9/18<br>Return: -1.0351% | Buy: 24.85; 2013/10/30<br>Sell: 25.04; 2013/12/6<br>Return: 0.76459% |
| 16 | | | | Buy: 68.91; 2013/9/19<br>Sell: 68.47; 2013/9/23<br>Return: -0.63851% | |
| 17 | | | | Buy: 67.39; 2013/10/15<br>Sell: 64.85; 2013/10/28<br>Return: -3.7691% | |
| Accumulated Return | 6.3% | 9.8% | 0% | 0.1% | 13% |
| Buy&Hold Return | 6.3% | -17.8% | -21.6% | 2.5% | 7.8% |

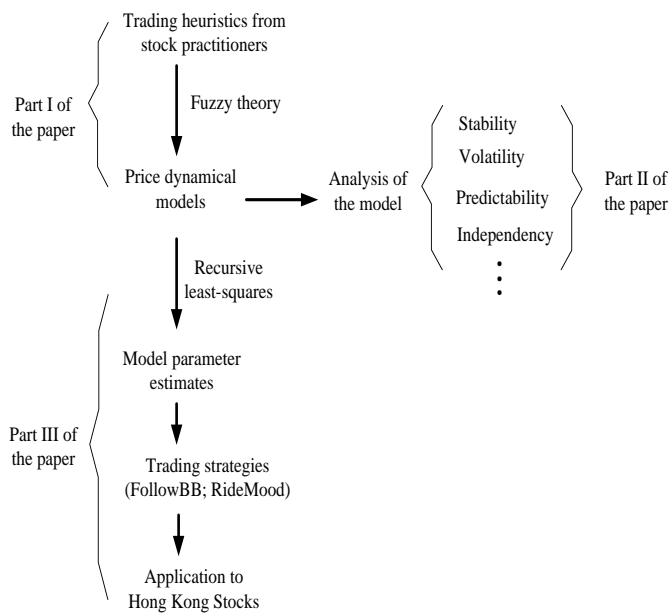

Fig. 15: The skeleton of the whole paper.

Technical trading rules are linguistic descriptions of trading wisdoms, and fuzzy systems theory provides good tools to translate them into the language of mathematics. In this paper we transformed only 12 technical trading heuristics into mathematics which cover only a very small portion of the dynamic field of technical analysis [24], [34], [39] and much more technical trading rules are waiting there to be transformed into mathematical formulas. For example, the whole branch of pattern rules [31] in technical analysis (e.g., price patterns with the names [24]: box, triangle, wedge, diamond, coil, megaphone, funnel, climax, saucer, bowl, cup, head-and-shoulder, flag, pennant, half-mast, oops, shark, doji, harami, hammer, hanging, shooting star, candle engulfing, dark cloud cover, piercing line, …) are not touched in this paper.

b. Analysis of more price dynamical models constructed from the technical trading rules. In Part II of this paper we analyzed only one price dynamical model --- the model based on the moving-average rules (Heuristic 1 in Part I of the paper), and it is clearly an important research topic to analyze the other technical-trading-rule-based price dynamical models (e.g., the models constructed from Heuristics 2 to 12 in Part I of the paper) in a mathematically rigorous fashion similar to what we did in Part II of this paper. For example, we saw in Part I of the paper that a lot of price jumps occur when the support, resistance, or trend line rules are activated. Can we find suitable probability distribution functions for these price jumps? How are these jumps related to the model parameters?

c. Using other optimization methods to estimate the model parameters. In Part III of this paper we used only the standard recursive least-squares algorithm with exponential forgetting to estimate the strength parameters of the big buyers/sellers. Since fast and accurate estimation of these time-varying parameters is crucial to the success of the trading strategies, it is clearly an important research topic to try other (perhaps more advanced) optimization algorithms to estimate the model parameters. A slight improvement in speed and accuracy for estimating the parameters could result in a big increase in portfolio returns.

d. Applying the FollowBB and RideMood strategies to other stock markets and different types of stocks to see for what classes of stocks and in what situations the strategies perform the best[9].

ACKNOWLEDGMENT

The author would like thank the anonymous AE and the reviewers for their insightful comments that helped to improve the paper. The author thanks Bill Lupien and Terry Rickard for comments on this work from the practitioner's perspective.

---

[9] "Stay Hungry. Stay Foolish." -- Steve Jobs.

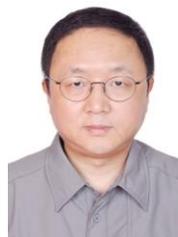

**Li-Xin Wang** received the Ph.D. degree from the Department of Electrical Engineering, University of Southern California (USC), Los Angeles, CA, USA, in 1992.

From 1992 to 1993, he was a Postdoctoral Fellow with the Department of Electrical Engineering and Computer Science, University of California at Berkeley. From 1993 to 2007, he was on the faculty of the Department of Electronic and Computer Engineering, The Hong Kong University of Science and Technology (HKUST). In 2007, he resigned from his tenured position at HKUST to become an independent researcher and investor in the stock and real estate markets in Hong Kong and China. In Fall 2013, he returned to academia and joined the faculty of the Department of Automation Science and Technology, Xian Jiaotong University, Xian, China, after a fruitful hunting journey across the wild land of investment to achieve financial freedom. His research interests are dynamical models of asset prices, market microstructure, trading strategies, fuzzy systems, and opinion networks.

Dr. Wang received USC's Phi Kappa Phi highest Student Recognition Award.